\documentclass[10pt,technote]{article}
\usepackage[cp1250]{inputenc}
\usepackage{amsmath}
\usepackage{pst-plot}
\usepackage{pstricks}
\usepackage{pst-node}
\usepackage{amssymb}
\usepackage{listings}
\usepackage{url}

\newtheorem{myDefi}{Definition}[subsection]
\newtheorem{lemi}[myDefi]{Lemma}
\newcommand{\proof}{\textbf{Proof:} }

\title{Linear Time Algorithms Based on Multilevel Prefix Tree for Finding 
Shortest Path with Positive Weights and Minimum Spanning Tree in a Networks
 \footnote{Cornell University Computing and Information Science 
           Technical Reports~\cite{TR}}
}
\author{David S. P\l aneta\footnote{dplaneta@gmail.com}}

\begin{document}
\maketitle
\begin{flushleft}
\begin{footnotesize}
\textbf{Categories and Subject Descriptors\footnote{The ACM Computing Classification System}:}\\
\texttt{C.2.1} [Computer-Communication Network]: \textsl{Network Architecture and Design--Distributed networks}\\
\texttt{E.1} [Data Structures]: \textsl{Graphs and networks}\\
\texttt{F.2.2} [Analysis of Algorithms and Problem Complexity]: \textsl{Nonnumerical Algorithms and Problems}\\
\texttt{G.2.2} [Discrete Mathematics]: \textsl{Graph Theory--trees}\\
\vspace{2ex}
\textbf{General Terms:} \textsl{Algorithms, Networks}\\
\vspace{2ex}
\textbf{Additional Key Words and Phrases:} 
\textsl{Minimal Spanning Tree, the Shortest Path, Single-Source Shortest 
Path, Single-Destination Shortest Path, MST, SSSP, SDSP}
\end{footnotesize}
\end{flushleft}

\addtolength{\textwidth}{-30pt}                                            %

\begin{abstract}
In this paper I present general outlook on questions relevant to the 
basic graph algorithms; Finding the Shortest Path with Positive Weights and 
Minimum Spanning Tree. 
I will show so far known solution set of basic graph problems and present my 
own. My solutions to graph problems  are characterized by their linear 
worst-case time complexity. It should be noticed that the algorithms which 
compute the Shortest Path and Minimum Spanning Tree problems not only 
analyze the weight of arcs (which is the main and often the only criterion 
of solution hitherto known algorithms) but also in case of identical path 
weights they select this path which walks through as few vertices as 
possible.
I have presented algorithms which use priority queue based on multilevel 
prefix tree -- PTrie. PTrie is a clever combination of the idea of prefix 
tree -- Trie, the structure of logarithmic time complexity for insert and 
remove operations, doubly linked list and queues. 
In C++ I will implement linear worst-case time algorithm computing the 
Single-Destination Shortest-Paths problem and I will explain its usage.
\break
\end{abstract}
\newpage
\section{Introduction}
\begin{figure}[!hbp]
\caption{Difference between MST and SSSP problems\label{DifferenceMSTvsSSSP}}
\begin{center}
\begin{pspicture*}(0,0)(12,5.4)

\rput[bc](2.5, 5.2){ \small{\textbf{Minimum Spanning Tree}} }
\rput[bc](9.5, 5.2){ \small{\textbf{Single-Source Shortest Path}} }

\ttfamily

\rput[bc](2.5, 4.5){\ovalnode{NodeA}{\textbf{A}}}
\rput[bc](1, 3.5){\ovalnode{NodeB}{\textbf{B}}}
\rput[bc](4, 3.5){\ovalnode{NodeC}{\textbf{C}}}
\rput[bc](2.5, 2.5){\ovalnode{NodeD}{\textbf{D}}}
\rput[bc](0.5, 1.5){\ovalnode{NodeE}{\textbf{E}}}
\rput[bc](4.5, 1.5){\ovalnode{NodeF}{\textbf{F}}}
\rput[bc](2.5, 0.5){\ovalnode{NodeG}{\textbf{G}}}

\ncline[doubleline=true, doublecolor=red]{-}{NodeB}{NodeA}\Aput{$1$}
\ncline{-}{NodeA}{NodeC}\Aput{$4$}
\ncline{-}{NodeA}{NodeD}\Aput{$3$}
\ncline{-}{NodeC}{NodeD}\Aput{$6$}
\ncline[doubleline=true, doublecolor=red]{-}{NodeC}{NodeF}\Aput{$1$}
\ncline{-}{NodeD}{NodeB}\Aput{$4$}
\ncline[doubleline=true, doublecolor=red]{-}{NodeD}{NodeE}\Aput{$2$}
\ncline[doubleline=true, doublecolor=red]{-}{NodeE}{NodeB}\Aput{$2$}
\ncline[doubleline=true, doublecolor=red]{-}{NodeF}{NodeD}\Aput{$1$}
\ncline[doubleline=true, doublecolor=red]{-}{NodeF}{NodeG}\Aput{$1$}
\ncline{-}{NodeG}{NodeE}\Aput{$5$}

\rput[bc](9.5, 4.5){\ovalnode[doubleline=true, doublecolor=red]{NodeA}{\textbf{A}}}
\rput[bc](8, 3.5){\ovalnode{NodeB}{\textbf{B}}}
\rput[bc](11, 3.5){\ovalnode{NodeC}{\textbf{C}}}
\rput[bc](9.5, 2.5){\ovalnode{NodeD}{\textbf{D}}}
\rput[bc](7.5, 1.5){\ovalnode{NodeE}{\textbf{E}}}
\rput[bc](11.5, 1.5){\ovalnode{NodeF}{\textbf{F}}}
\rput[bc](9.5, 0.5){\ovalnode{NodeG}{\textbf{G}}}

\ncline[doubleline=true, doublecolor=red]{-}{NodeB}{NodeA}\Aput{$1$}
\ncline[doubleline=true, doublecolor=red]{-}{NodeA}{NodeC}\Aput{$4$}
\ncline[doubleline=true, doublecolor=red]{-}{NodeA}{NodeD}\Aput{$3$}
\ncline{-}{NodeC}{NodeD}\Aput{$6$}
\ncline{-}{NodeC}{NodeF}\Aput{$1$}
\ncline{-}{NodeD}{NodeB}\Aput{$4$}
\ncline{-}{NodeD}{NodeE}\Aput{$2$}
\ncline[doubleline=true, doublecolor=red]{-}{NodeE}{NodeB}\Aput{$2$}
\ncline[doubleline=true, doublecolor=red]{-}{NodeF}{NodeD}\Aput{$1$}
\ncline[doubleline=true, doublecolor=red]{-}{NodeF}{NodeG}\Aput{$1$}
\ncline{-}{NodeG}{NodeE}\Aput{$5$}

\end{pspicture*} 
\end{center}
\end{figure}
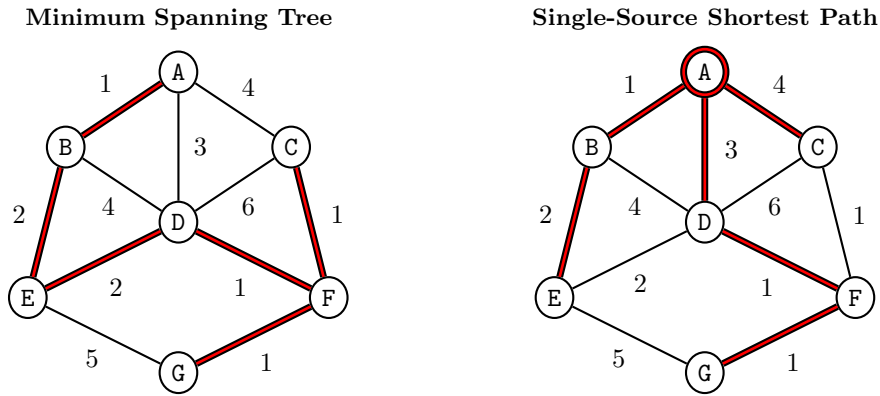

Graphs are a pervasive data structure in computer science and algorithms 
for working with them are fundamental to the field. There are hundreds of 
interesting computational problems defined in terms of graph. 
A lot of really complex processes can be solved in a very effective and 
clear way by means of terms of graph. Algorithms which solve graph problems 
are implemented in many appliances of everyday use. They help flight control 
system to administer the airspace. They are crucial for economists to do 
market research, mathematicians to solve complicated problems. And finally, 
they help programmers describe object connections. Every day, many people 
trust graph algorithms when they, implemented in GPS system, calculate the 
shortest way to their destination. There are many basic graph algorithms, 
whose computational complexity is of greatest importance. 
They include algorithms on directed graphs finding Single-Source Shortest 
Path with positive weights~(SSSP) and Minimum Spanning Tree~(MST) 
[Figure~\ref{DifferenceMSTvsSSSP}]. 
Based on multilevel prefix tree (PTrie~\cite{Planeta06}) I compute these 
problems in linear worst-case time and in case of identical path weights 
it selects those paths which walk through as few vertices as possible.
\break
%
\subsection{Previous work about MST}
Algorithm computing MST problem is frequently used by administrators, who 
think how to construct the framework of their networks to connect all 
servers in which they use as little optical fiber as possible. Not only 
computer engineers use algorithms based on MST. Architecture, electronics 
and many other different areas take advantage of algorithms using MST.
The MST problem is one of  the oldest and most basic graph problems in 
computer science. The first MST algorithm was discovered by 
Bor\r uvka~\cite{Boruvka26} in 1926 
(see~\cite{Nesetril01} for an English translation). 
In fact, MST is perhaps the oldest open problem in computer 
science. Kruskal's algorithm was reported by Kruskal~\cite{Kruskal56} 
in 1956. The algorithm commonly known as Prim's algorithm was indeed 
invented by Prim~\cite{Prim57} in 1957, but it was also invented earlier by 
Vojtech Jarn\'ik in 1930. Effective notation of these algorithms require 
$O(|V|log|E|)$ time, where $|V|$ and $|E|$ denote, respectively, the number 
of vertices and edges in the graph. In 1975, Yao~\cite{Yao75} first 
improved MST to $O(|E|loglog|V|)$, which starts with all nodes as fragments, 
extends each fragment, then combines, then extends each of the new enlarged 
fragments, then combines again, and so forth. In 1985, using a combination 
of the ideas from Prim's algorithm, Kruskal's algorithm and Bor\r uvka's 
algorithm, together, Fredman and Tarjan~\cite{Fredman87} give on algorithm 
that runs in $O(|E|\beta(|E|,|V|))$ using Fibonacci heaps, where 
$\beta(|E|,|V|)$ is the number of log iterations on $|V|$ needed to make it 
less than $\frac{|E|}{|V|}$. As an alternative to Fibonacci heaps we 
can use insignificantly improved Relaxed heaps~\cite{Driscoll88}.
Relaxed heaps also have some advantages over Fibonacci heaps in parallel 
algorithms.
Shortly after, Gabow, Galil, Spencer, and Tarjan~\cite{Gabow86} improved 
this algorithm to run in $O(|E|log\beta(|E|,|V|)$. 
In 1999, Chazelle~\cite{Chazelle00} takes a significant step towards a 
solution and charts out a new line of attack, gives on algorithm that 
runs in $O(|E|\alpha(|E|,|V|))$ time, where $\alpha(|E|,|V|)$ is the 
function inverse of Ackermann's function~\cite{Tarjan75}. 
Unlike previous algorithms, Chazelle's algorithm does not follow the greedy 
method. In 1994, Fredman and Willard~\cite{Fredman94} showed how to find 
a minimum spanning tree in $O(|V| + |E|)$ time using a deterministic 
algorithm that is not comparison based. Their algorithm assumes that the 
data are $b$-bit integers and that the computer memory consists of 
addressable $b$-bit words.\break

A great many of so far invented implementations attain linear time on 
average runs but their worst-case time complexity is higher. The invention 
of versatile and practice algorithm running in linear worst-case time 
still remains an open problem. For decades many researchers have been 
trying to do find linear worst-case time algorithm solving MST problem. 
To find this algorithm researchers start with Boruvka's algorithm and 
attempt to make it run in linear worst-case time by focusing on the 
structures used by the algorithm.
\subsection{Previous work about SSSP}
The Single-Source Shortest Path on directed graph with positive weight 
(SSSP) is one of the most basic graph problems in theoretical computer 
science. This problem is also one of the most natural network optimization 
problems and occurs widely in practice. SSSP problem consists in finding 
the shortest (minimum-weight) path from the source vertex to every other 
vertex in the graph. The Shortest Paths algorithms typically rely on the 
property that the shortest path between two vertices contains other shortest 
paths within it. Algorithms computing SSSP problem are used in considerable 
amount of applications. Starting from rocket software and finishing with 
GPS inside our cars. In many programs algorithm computing SSSP problem is a 
part of basic data analysis.
For example, itineraries, flight schedules and other transport systems can 
by presented as networks, in which various shortest path problems are very 
important. We many aim at making the time of flight between two cities as 
short as possible or at minimizing the costs. In such networks the casts 
may concern time, money or some other resources. In these networks 
particular resources don't have to be dependent. It should be noted that in 
reality price of the ticket may not be a simple function of the distance 
between two cities - it is quite common to travel cheaper by taking a 
roundabout route nether  than a direct one. Such difficulties can be 
overcome by means of algorithms solving the shortest path problems.
Algorithm computing SSSP problem are often used in real time systems, where 
is of great importance every second. Like OSPF (open shortest path first)
~\cite{OSPF} is a well known real-world implementation of SSSP algorithm 
used in internet routing. That's why time the efficiency of algorithms 
computing SSSP problem is very important. By reversing the direction of 
each edge in the graph, we receive Single-Destination Shortest-Paths 
problem (SDSP); the Shortest Path to a given destination source vertex 
from each vertex.\break

Dijkstra's algorithm was invented by Dijkstra~\cite{Dijkstra59} in 1959, 
but it contained no mention of priority queue, needs $O(|V|^2 + |E|)$ time. 
The running time of Dijkstra's algorithm depends on how the min-priority 
queue is implemented. If the graph is sufficiently sparse-in particular, 
$E = o(\frac{|V|^2}{lg|V|})$ - it is practical to implement the min-priority 
queue with a binary min-heap. Then the time of the algorithm~\cite{Johnson77} 
is $O(|E|log|V|)$. In fact, we can achieve a running time of $O(|V|lg|V| + |E|)$ 
by implementing the min-priority queue with Fibonacci heap~\cite{Fredman87}. 
Historically, the development of Fibonacci heaps was motivated by the 
observation that in Dijkstra's algorithm there are, typically, many more 
decrease-key calls than extract-min calls, so any method of reducing the 
amortized time of each decrease-key operation to $o(lg|V|)$ without 
increasing the amortized time of extract-min would yield on asymptotically 
faster implementation than with binary heaps. But Goldberg and 
Tarjan~\cite{Goldberg96} observed in practice and helped to explain why 
Dijkstra's codes based on binary heaps perform better than the ones based on 
Fibonacci heaps. A number of faster algorithms have been developed
on more powerful RAM (random access machine) model. In 1990, Ahuja, 
Mehlhorn, Orlin, and Tarjan~\cite{Ahuja90} give on algorithm that runs in 
$O(|E| + |V|\sqrt{logW})$, where $W$ is the largest weight of any edge in 
graph. In 2000, Thorup~\cite{Thorup00} gives on 
$O(|V| + |E|loglog|V|)$ time algorithm. Faster approaches for 
somewhat denser graphs have been proposed by Raman~\cite{Raman97} in 1996. 
Raman's algorithm require $O(|E| + |V|\sqrt{log|V|loglog|V|})$ and 
$O(|E| + |V|(W * log|V|)^{1/3})$ time, respectively. However
Asano~\cite{Asano00} shows, the algorithms don't perform well in practical 
simulations.  The classic label-correcting algorithm of Bellman-Ford is 
based on separate algorithms by Bellman~\cite{Bellman58}, published in 1958, 
and Ford~\cite{Ford62}, published in 1956~\cite{Ford56}, and all of its improved 
derivatives~\cite{Cherkassky96}\cite{Deo84}\cite{Pollack60}\cite{Glover84}%
\cite{Bertsekas93} need $\Omega(|V|*|E|)$ time in worst time. 
However in case of graph with irrationally heavy weight of edges algorithm's 
may possibly equal $O(2^{|V|})$ time cost~\cite{Galo86}.
But Bellman-Ford algorithm not only computes the single-source shortest path 
with positive weights, but also solves the single-source shortest path 
problem in the general case in which edge weights may be 
negative. The Bellman-Ford algorithm returns a boolean value indicating 
whether or not there is a negative-weight cycle that is reachable from the 
source. If there is such a cycle, the algorithm indicates that no solution 
exists. If there is no such cycle, the algorithm produces the Shortest 
Paths and their weights.\break
\section{Linear worst-case time algorithms based on multilevel prefix tree 
         computing the basic network problems}
I show algorithms which use priority queue based on multilevel prefix tree 
-- PTrie~\cite{Planeta06}. PTrie is a clever combination of the idea of 
prefix tree - Trie~\cite{Briandais59}, the structure of logarithmic time 
complexity for insert and remove operations, doubly linked list~\cite{Knuth1} 
and queues~\cite{Knuth1}.\\

I assume that algorithms which I present the weight of edges is constant. 
The weight of edges are in $\{0, …, 2^t - 1\}$, where $t$ denotes the word 
length (size of). For all edges of graph $G=(V,E)$ the constant value 
$t_{max}$ can be matched. In other words, I assume that the size of type 
which remembers the  weight of edges is constant and identical for all 
edges of graph $G=(V,E)$.
\subsection{PTrie: Priority queue based on multilevel prefix tree}
Priority Trie (PTrie) uses a few structures including Trie of 
$2^K$ degree~\cite{Briandais59}, which is the structure core. Data recording 
in PTrie consists in breaking the word into parts which make the indexes of 
the following layers in the structure (table look-at). The last layers 
contain the addresses of doubly linked list's nodes. Each of the list nodes 
stores the queue~\cite{Knuth1}, into which the elements are inserted. 
Moreover, each layer contains the structure of logarithmic time complexity 
of insert and remove operations. Which help to define the destination of data 
in the doubly linked list~\cite{Knuth1}.
%
%
\subsubsection{Terminology}
Bit pattern is a set of $K$ bits. $K$ (length of bit pattern) defines the 
number of bits which are cut off the binary word. 
$M$ defines number (length) of bits in a binary word.
\begin{center}
value of word = $\overbrace{\underbrace{101...1}_{K}00101...}^{M}$
\end{center}
$N$ is  number of all values of PTrie. $2^K$ is variation $K$ of element 
binary set \{$0,1$\}. It determines the number of groups 
(number of Layers [Figure~\ref{Layer}]), which the bit pattern may be 
divided into during one step (one level). 
The set of values decomposed into the group by the first $K$
bits (the version of algorithm described in paper was implemented by 
machine of little-endian type).
\begin{figure}[!hbp]
\caption{Layer\label{Layer}}
\begin{center}
\begin{pspicture*}(0,4)(9,9.2)

\scriptsize
\rput[bc](6,8.4){\ovalnode{NodeA}{\textbf{A}}}
\rput[bc](7,7.4){\ovalnode{NodeC}{\textbf{C}}}
\rput[bc](5,7.4){\ovalnode{NodeB}{\textbf{B}}}
\rput[bc](4,6.4){\ovalnode{NodeD}{\textbf{D}}}
\ncline{<-}{NodeB}{NodeA}
\ncline{<-}{NodeC}{NodeA}
\ncline{<-}{NodeD}{NodeB}

\footnotesize
\rput[bl](6,6.5){ \textbf{\normalsize{\ldots}} }

\psline{|-|}(7.6,9)(7.6,6)
\rput[bl](7.8,6){\rotateright{$log_{2}P = log_{2}2^K = K$}}

\pscurve(3.6,6)(3.5,6)(3.2,6.4)(3.2,7)(3,7.5)%
        (3,7.5)(3.2,8)(3.2,8.6)(3.5,9)(3.6,9)
\rput[bl](0.2,6.8){\parbox{2.7cm}{The Structure of logarithmics time complexity of insert and remove operations}}

 \rput[bl](0.2,5){\rnode{BlockMIN}{\psdblframebox{ \parbox{18pt}{MIN} }}}
 \rput[bl](1.6,5){\rnode{BlockMAX}{\psdblframebox{ \parbox{18pt}{MAX} }}}
 
 \rput[bl](3,5){\rnode{blockA}{\psframebox{ $00....00 $ }}}
 \rput[bl](3.6,5.6){$G_{1}$}
 
 \rput[bl](4.8,5){\rnode{blockB}{\psframebox{ $ 00...01 $ }}}
 \rput[bl](5.4,5.6){$G_{2}$}
 
 \rput[bl](6.2,5){ \textbf{\normalsize{\ldots}} }
 
 \rput[bl](7,5){\rnode{blockC}{\psframebox{ $ 11...11 $ }}}
 \rput[bl](7.6,5.6){$G_{P}$}
 
 \psline{|-|}(2.9,4.6)(8.4,4.6)
 \rput[bc](5.6, 4.4) {$P=2^K$}
 
\end{pspicture*}
\end{center}
\end{figure}
The path is defined starting from the most important bits of variable. 
The value of pattern $K$ (index) determines the layer we move to 
[Figure~\ref{PTrie}]. The lowest layers determine the nodes of the 
list which store the queues for inserted values. L defines the level 
the layer is on. 
\begin{figure}[!hbp]
\begin{center}
\caption{PTrie\label{PTrie}}
\begin{pspicture*}(-1,-1)(11,9.4)
\scriptsize
 \rput[bc](6,8.4){\ovalnode{NodeA}{\textbf{A}}}
 \rput[bc](7,7.4){\ovalnode{NodeC}{\textbf{C}}}
 \rput[bc](5,7.4){\ovalnode{NodeB}{\textbf{B}}}
 \rput[bc](4,6.4){\ovalnode{NodeD}{\textbf{D}}}
 \ncline{<-}{NodeB}{NodeA}
 \ncline{<-}{NodeC}{NodeA}
 \ncline{<-}{NodeD}{NodeB}

\footnotesize
 \rput[bl](6,6.5){ \textbf{\normalsize{\ldots}} }

 \psline{|-|}(7.6,9)(7.6,6)
 \rput[bl](7.8,6.4){\rotateright{$\Theta(log_{2}2^K) = O(K)$}}

 \rput[bl](0.2,8.8){\textbf{Layer}}

 \rput[bl](0.2,5){\rnode{BlockMIN}{\psdblframebox{ \parbox{18pt}{MIN} }}}
 \rput[bl](1.6,5){\rnode{BlockMAX}{\psdblframebox{ \parbox{18pt}{MAX} }}}
 
 \rput[bl](3,5){\rnode{BlockA}{\psframebox{ $00....00 $ }}}
 \rput[bl](3.6,5.6){$G_{1}$}
 
 \rput[bl](4.8,5){\rnode{BlockB}{\psframebox{ $ 00...01 $ }}}
 \rput[bl](5.4,5.6){$G_{2}$}
 
 \rput[bl](6.2,5){ \rnode{Block...}{\textbf{\normalsize{\ldots}} }}
 
 \rput[bl](7,5){\rnode{BlockC}{\psframebox{ $ 11...11 $ }}}
 \rput[bl](7.6,5.6){$G_{P}$}
 
 \psline{|-|}(2.9,4.6)(8.4,4.6)
 \rput[bc](5.6, 4.4) {$P=2^K$}
 
 \psframe[dimen=inner](0,4)(9,9.2)

\psframe[dimen=inner](0,2)(2,3)
\rput(1,2.5){\rnode{LayerA}{\textbf{Layer}}}
\ncline{<-}{LayerA}{BlockA}

\psframe[dimen=inner](3,2)(5,3)
\rput(4,2.5){\rnode{LayerB}{\textbf{Layer}}}
\ncline{<-}{LayerB}{BlockB}

\psframe[dimen=inner](8,2)(10,3)
\rput(9,2.5){\rnode{LayerC}{\textbf{Layer}}}
\ncline{<-}{LayerC}{BlockC}

\rput[bl](6.2,2.2){ \rnode{Layer...}{\textbf{\normalsize{\ldots}} }}
\ncline{<-}{Layer...}{Block...}

\rput[bl](0.5,1.5){ \rnode{dots1}{\textbf{\normalsize{\ldots}} }}
\rput[bl](2,1.5){ \rnode{dots2}{\textbf{\normalsize{\ldots}} }}
\rput[bl](3,0.2){ \rnode{dots3}{\textbf{\normalsize{\ldots}} }}
\rput[bl](3,1.5){ \rnode{dots4}{\textbf{\normalsize{\ldots}} }}
\rput[bl](4.1,1.5){ \rnode{dots5}{\textbf{\normalsize{\ldots}} }}
\rput[bl](6.2,0.2){ \rnode{dots6}{\textbf{\normalsize{\ldots}} }}
\rput[bl](8.6,1.5){ \rnode{dots7}{\textbf{\normalsize{\ldots}} }}

\ncline{<-}{dots1}{LayerA}
\ncline{<-}{dots2}{LayerA}
\ncline{<-}{dots3}{dots2}
\ncline{<-}{dots4}{LayerB}
\ncline{<-}{dots3}{dots4}
\ncline{<-}{dots5}{LayerB}
\ncline{<-}{dots6}{Layer...}
\ncline{<-}{dots7}{LayerC}

\psline{|-|}(10.2,9.4)(10.2,1)
\rput[bl](10.3,3){\rotateright{$\Theta(log_{2^K}N) = \Theta(\frac{lgN}{lg2^K}) = O(\frac{M}{K})$}}
 
\rput[bl](-0.8,6.5){$L_{1}$}
\rput[bl](-0.8,2.5){$L_{2}$}
\rput[bl](-0.8,1.4){$L_{log_{2^K}N}$}
\rput[bl](0,0){\rnode{Node1}{\psframebox{$Node$}}}
\ncline{<-}{Node1}{dots1}
\rput[bl](0,-1){\rnode{Q1}{\psframebox{\scriptsize{Queue}}}}
\ncline{<-}{Q1}{Node1}

\rput[bl](1.5,0){\rnode{Node2}{\psframebox{$Node$}}}
\ncline{<-}{Node2}{dots1}
\rput[bl](1.5,-1){\rnode{Q2}{\psframebox{\scriptsize{Queue}}}}
\ncline{<-}{Q2}{Node2}

\rput[bl](4.2,0){\rnode{Node3}{\psframebox{$Node$}}}
\ncline{<-}{Node3}{dots5}
\rput[bl](4.2,-1){\rnode{Q3}{\psframebox{\scriptsize{Queue}}}}
\ncline{<-}{Q3}{Node3}

\rput[bl](7.5,0){\rnode{Node4}{\psframebox{$Node$}}}
\ncline{<-}{Node4}{dots7}
\rput[bl](7.5,-1){\rnode{Q4}{\psframebox{\scriptsize{Queue}}}}
\ncline{<-}{Q4}{Node4}

\rput[bl](9,0){\rnode{Node5}{\psframebox{$Node$}}}
\ncline{<-}{Node5}{dots7}
\rput[bl](9,-1){\rnode{Q5}{\psframebox{\scriptsize{Queue}}}}
\ncline{<-}{Q5}{Node5}

\rput[tl](10.5,1){\rnode{Tail}{\rotateright{\psframebox{Tail}} }}
\nccurve[angleB=270]{<-}{Node5}{Tail}

\rput[tl](-1,1){\rnode{Head}{\rotateright{\psframebox{Head}} }}
\nccurve[angleB=180, angleA=270]{->}{Head}{Node1}

\ncarc{->}{Node1}{Node2}
\ncarc{->}{Node2}{Node1}

\ncarc{->}{dots3}{Node2}
\ncarc{->}{Node2}{dots3}

\ncarc{->}{dots3}{Node3}
\ncarc{->}{Node3}{dots3}

\ncarc{->}{dots6}{Node3}
\ncarc{->}{Node3}{dots6}

\ncarc{->}{dots6}{Node4}
\ncarc{->}{Node4}{dots6}

\ncarc{->}{Node5}{Node4}
\ncarc{->}{Node4}{Node5}

\end{pspicture*}
\end{center}
\end{figure}
Probability that exactly $G$ keys correspond to one particular pattern, 
where for each of $P_{L}$ sequences of leading bits there is such a node 
that corresponds to at least two keys equals
\begin{center}
 ${N \choose G} P^{-GL}(1-P^{-L})^{N-G}$
\end{center}
For random PTrie the average number of layers on level $L$, for 
$L = 0, 1, 2, \ldots $is
\begin{center}
 $P^L(1-(1-P^{-L})^N)-N(1-P^{-L})^{N-1}$
\end{center}
If $A_{N}$ is average number of layers in random PTrie of degree $P=2^K$ 
containing $N$ keys. Then $A_{0} = A_{1} = 0$, and for $N \geq 2$ 
we get~\cite{Knuth3}:
\begin{center}
 \begin{displaymath}
  A_{N} = 1 + \sum_{G_{1} + \ldots + G_{P} = N} 
  \Big(\frac{N!}{G_{1}! \ldots G_{P}!} P^{-N}\Big) 
  \Big(A_{G_{1}} + \ldots + A_{G_{P}}\Big) = 
 \end{displaymath}
 \begin{displaymath}
  1 + P^{1-N} \sum_{G_{1} + \ldots + G_{P} = N}
  \Big(\frac{N!}{G_{1}! \ldots G_{P}!}\Big) A_{G_{1}} =
 \end{displaymath}
 \begin{displaymath}
  1 + P^{1-N} \sum_{G}{N \choose G} \Big(P - 1 \Big)^{N-G} A_{G} =
 \end{displaymath}
  \begin{displaymath}
  1 + 2^{G(1-N)} \sum_{G} {N \choose G} \Big( 2^G - 1 \Big)^{N-G} A_{G}
 \end{displaymath}
\end{center}
%
%
\subsubsection{Implementation}
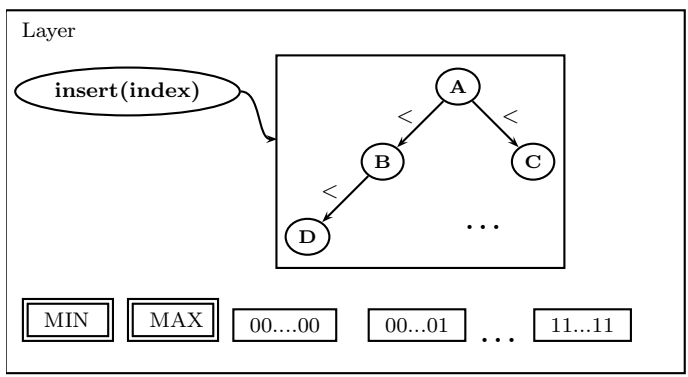
\begin{figure}[!hbp]
\caption{Insert value of index into the ordered binary tree from the layer\label{Tree}}
\begin{center}
\begin{pspicture*}(0,4.5)(9.2,9.5)

\scriptsize
\rput[bc](6,8.4){\ovalnode{NodeA}{\textbf{A}}}
\rput[bc](7,7.4){\ovalnode{NodeC}{\textbf{C}}}
\rput[bc](5,7.4){\ovalnode{NodeB}{\textbf{B}}}
\rput[bc](4,6.4){\ovalnode{NodeD}{\textbf{D}}}
\ncline{<-}{NodeB}{NodeA}
\ncline{<-}{NodeC}{NodeA}
\ncline{<-}{NodeD}{NodeB}

\footnotesize
\psframe[dimen=inner](3.6,6)(7.4,8.8)
\rput[bl](3.6,7.7){\rnode{square}{}}
\rput[bl](0.2,9){Layer}
\rput[bl](0.1,8){\ovalnode{insert}{\textbf{insert(index)}}}
\nccurve[angleB=180]{->}{insert}{square}

\rput[bc](5.3,8){ \textbf{$<$} }
\rput[bc](6.7,8){ \textbf{$<$} }
\rput[bc](4.3,7){ \textbf{$<$} }

\rput[bl](6,6.5){ \textbf{\normalsize{\ldots}} }

 \rput[bl](0.2,5){\rnode{BlockMIN}{\psdblframebox{ \parbox{18pt}{MIN} }}}
 \rput[bl](1.6,5){\rnode{BlockMAX}{\psdblframebox{ \parbox{18pt}{MAX} }}}
 \rput[bl](3,5){\rnode{blockA}{\psframebox{ $00....00 $ }}}
 \rput[bl](4.8,5){\rnode{blockB}{\psframebox{ $ 00...01 $ }}}
 \rput[bl](6.2,5){ \textbf{\normalsize{\ldots}} }
 \rput[bl](7,5){\rnode{blockC}{\psframebox{ $ 11...11 $ }}}

\psframe[dimen=inner](0,4.6)(9,9.4)
\end{pspicture*}
\end{center}
\end{figure}

\begin{center}
\begin{tabular}{|l|l|l|}
\hline
\textbf{Operation} & \textbf{Description} & \textbf{Bound}\\
\hline
create & Creates object & $O(1)$ \\
\hline
insert(data) & Adds element to the structure. & $O(\frac{M}{K} + K)$\\
\hline
boolean remove(data) & 
\parbox{5.5cm}{Removes value from the tree. If operation failed because there 
was no such value in the tree it returns FALSE(0), otherwise returns TRUE(0).}
& $O(\frac{M}{K} + K)$\\
\hline
boolean search(data) &
\parbox{5.5cm}{
Looks for the words in the tree. If finds return TRUE(1), 
otherwise FALSE(0).}
& $O(\frac{M}{K})$\\
\hline
*minimum() & 
\parbox{5.5cm}{
Returns the address of the lowest value in the tree, or empty address if the 
operation failed because the tree was empty.
}
& $O(1)$\\
\hline
*maximum() & 
\parbox{5.5cm}{
Returns the address of the highest value in the tree or empty address if the 
operation failed because the tree was empty.
}
& $O(1)$\\
\hline
next & 
\parbox{5.5cm}{
Returns the address of the next node in the tree or empty address if value 
transmitted in parameter was the greatest. The order of moving to successive 
elements is fixed - from the smallest to the largest and from 
``the youngest to the oldest'' (stable) in case of identical words.
}
& $O(1)$\\
\hline
prev &
\parbox{5.5cm}{
Similar to `next' but it returns the address of preceding node in the tree.
}
& $O(1)$\\
\hline
\end{tabular}
\end{center}
Basic operations can be joined. For example, the effect connected with the 
heap; delete-min() can be replaced by operations remove(minimum()).
\\
\\
\textbf{Insert}\\
Determine the interlinked index (pointer) to another layer using the length 
of pattern projecting on the word.\\
\indent\emph{\textbf{If}} interlink determined by index is not empty and indicated 
the list node -- try to insert the value into the queue of determined node.\\
\emph{If} the elements in the queue turn out to be the same,  
insert value into the queue.
\emph{Otherwise}, if elements in the queue are different from 
the inserted value, the node is ``pushed'' to a lower level and the hitherto 
existing level (the place of node) is complemented with a new layer. 
Next, try again to insert the element, this time however, into the newly 
created layer.\\
\indent\emph{\textbf{Else}}, if the interlink determined by index is empty, insert 
value of index into the ordered binary tree from the current layer 
[Figure~\ref{Tree}]. Father of a newly created node in ordered binary 
tree from the current layer determines the place for leaves; If the newly 
created node in ordered binary tree is on the right side of father 
(added index $>$ father index), the value added to the list will be 
inserted after the node determined by father index and the path of the 
highest indexes (make use of pointer `max' of the layers -- time cost $O(1)$) 
of lower level layers. If newly created node is on the left side of father 
(added index $<$ father index), the value added to the list will be 
inserted before the node determined by father index and the path of the 
smallest indexes (make use of pointer `min' of the layers -- time cost $O(1)$) 
of lower level layers.\break
\noindent%
One can wonder why we use the queue and not the stack or the value counter. 
Value counter cannot be used because complex elements can be inserted into 
PTrie structure, distinguishable in the tree only because of some words. 
Also, it is not a good idea to use a stack because the queue makes the 
structure stable. And this is a very useful characteristic.
I used ``plain'' Binary Search Tree in the structure of logarithmic time 
complexity. For a small number of tree nodes it is a very good solution 
because for $K = 4$, $2^K = 16$. So in the tree there may be maximum $16$ 
(different) elements. For such a small amount of (different) values the 
remaining ordered trees will probably turn out to be at most as effective 
as unusually simple Binary Search Trees.\\
%
%
\textbf{Analysis:}
In case of random data it will take 
$\Theta(\frac{lgN}{lg2^K}) = \Theta(log_{2^K}N) = O(\frac{M}{K})$ goings 
through layers to find the place in the heap core -- Trie tree. 
On at least one layer of PTrie structure we will use inserting into the 
ordered binary tree in which maximum number of nodes is $2^K$. 
While inserting the new value I need information where exactly it will be 
located in the list. Such information can be obtained in two ways; 
I will get the information if the representation of the nearest index on the 
list is to the left or to the right side of the inserted word index. 
It may happen that in the structure there is already is exactly the same 
word as the inserted one. In such case value index won't be inserted into 
any layer of the PTrie because it will not be necessary to add a new 
node of the list. Value will be inserted into the queue of already 
existing node. To sum up, while moving through the layers of PTrie we can 
stop at some level because of empty index. Then, a node will be added to 
the list in place determined by binary search tree and the remaining 
part of the path. This is why the bound of operation which inserts new 
value into PTrie equals
$\Theta(log_{2^K}N +log_{2}{2^K}) = \Theta(log_{2^K}N + K) = O(\frac{M}{K} + K)$.
\break
%
%
\\
\textbf{Find}\\
Method find like in case of plain Trie trees goes through succeeding layers 
following the path determined by binary representation of search value. 
It can be stated that it uses number key as a guide while moving down the 
core of PTrie -- prefix tree. In case of searching tree things can happen:
\begin{itemize}
	\item We don't reach the node of the list because the index we determine 
	      is empty on any of layers -- searching failure.
	\item We reach the node but values from the queue are different from the 
	      searched value -- searching failure.
	\item We reach the node and the values from the queue are exactly like 
	      the ones we seek -- searching success.
\end{itemize}
%
%
\textbf{Analysis:}
Searching in prefix tree is very fast because it finds the words using word 
key as indexes. In case of search failure the longest match of a searched 
word is found. It must be taken into consideration that during operation 
`search' we use only the attributes of prefix tree. This is why the amount 
of search numbers looked through during the random search is 
$\Theta(log_{2^K}N) = O(\frac{M}{K})$.\\
%
%
\\
\textbf{Remove}\\
Remove method just like find method ``moves down'' the PTrie structure to 
seek for the element to be deleted. If it doesn't reach the node of 
the list, or it does but the search value is different from the value of 
node queue, it does not delete any element of PTrie because it is not there. 
However if it reaches the node of the list and search value turns out to be 
the value from the queue -- it removes the value from the queue. 
If it remains empty after removing the element from the queue the node 
will be removed from the list and will return to the ``upper'' layers 
of prefix tree to delete possible, remaining, empty layers.\\
%
%
\textbf{Analysis:}
Since it is possible not only to go down the tree but also come back upwards 
(in case of deleting of the lower layer or the node of the list) the total 
length of the path move on is limited $\Theta(2log_{2^K}N)$. If delete 
the layer, it means there was only one way down from that layer, which 
implicates the fact that the ordered binary tree of a given layer contained 
only one node (index). The layer is removed if it remains empty after 
the removal of node from ordered binary tree. So the number of operation 
necessary for the removal of the layer containing one element equals 
$\Theta(1)$. In case of removal of layer $L_{i}$, if ordered binary tree 
of higher level layer $L_{i-1}$, despite removing the node which determines 
empty layer we came from, does not remain empty it means that there could be 
maximum $2^K$ nodes in the ordered binary tree. Operation of value delete 
from ordered binary tree amounts to $\Theta(log_{2}2^K) = \Theta(K)$. 
There is no point of ``climbing'' up the upper layers, since the layer we 
came from would not be empty. At this stage the method remove ends. 
To sum up, worse time complexity of remove operation is 
$\Theta(2log_{2^K}N + K) = O(\frac{M}{K} + K)$.\break
%
%
\\
\textbf{Extract minimum and maximum}\\
If the list is not empty, `minimum' reads the value pointed by the head of 
the list and `Maximum' reads the value pointed by the tail of the list.\\
%
%
\textbf{Analysis:}
Time complexity of operations is $\Theta(1)$.
%
%
\\
\\
\textbf{Iterators}\\
The nodes of the list are linked. If we know the position of one of the nodes, 
we have a direct access to its neighbors. The `next' operation reads the 
successor of current pointed node. The `prev' operation reads the predecessor 
of currently pointed node.\\
%
%
\textbf{Analysis:}
Moving to the node its neighbor requires only reading of the contents of the 
pointer `next' or `prev'. Time complexity of such operations equals 
$\Theta(1)$.
\subsubsection{Correctness}
PTrie has been designed like this, so as not to assume that keys have to be 
positive numbers or only integers - they can be even strings (however, 
in most cases the weight of arcs is represented by numbers).
To insert PTrie negative and positive integers I use not one PTrie, but two! 
One of the structures is destined exclusively for storing positive integers 
and the other one for storing only negative integers. The latter structure 
of PTrie is responsible only for negative integers - the integers are stored 
in reverse order on the list (for machine of little-endian type).
Therefore in case of the second structure of PTrie (responsible only for 
negative integers) I used standard operation of PTrie: PTrie2.maximum to 
extract the smallest value. Also real numbers (for example in 
ANSI IEEE 754-1985 standard~\cite{IEEE_754-1985}) can be used of the 
description of the weight of arcs on condition that two 
interrelated structures of PTrie will be used to put off exponent and 
mantissa. It is possible, because implementation of PTrie~\cite{Planeta06} 
described by me uses queue, which makes it stable. One of the structures 
of PTrie serves as storage for exponent, where each of the nodes of the 
list will contain additional structure of PTrie to store mantissa.
\subsubsection{Conclusion}
Efficiency of PTrie considerably depends on the length of pattern $K$. 
$K$ defines optional value, which is the power of two in the range 
[$1$, min($M$)]. The total size of necessary memory bound is proportional to 
$\Theta(\frac{log_{2^K}N(2^{K+1})}{K})$ because the number of layers 
required to remember $N$ random elements in PTrie of degree $2^K$ equals 
$\frac{lgN}{lgP}*P$. Moreover, each layers has tree of maximum size $2^K$ 
nodes and table of the $P$-elements, so the necessary memory bound equal 
$\Theta(log_{2^K}N * 2P) = \Theta(\frac{M}{K} * 2^{K+1})$. 
For data types of constant size maximum Trie tree height equals 
$\frac{M}{K}$. So the pessimistic operation time complexity is 
$O(\frac{M}{K} + K)$. For example, for four-byte numbers it is the most 
effective to determine the pattern $K=4$ bits long. Then, the pessimistic 
number of steps necessary for the operation on the PTrie will equal 
$\Theta(\frac{M}{K} + K) = \frac{32}{4} + 4 = 12$. 
Increasing $K$ to $K=8$ does not increase the efficiency of the structure 
operation because $\Theta(\frac{M}{K} + K) = \frac{32}{8} + 8 = 12$. 
What is more, in will unnecessarily increase the memory demand. A single 
layer consisting of $P=2^K$ groups for $K=8$ will contain tables 
$P=2^8=256$ long, not when $K=4$, only $P=2^4=16$ links. For variable size 
data the time complexity equals $\Theta(log_{2^K}N + K)$. Moreover, the 
length of pattern $K$ must be carefully matched. For example, for strings $K$ 
should not be longer than $8$ bits because we could accidentally read the 
contents from beyond the string which normally consist of one-byte sign! 
It is possible to record data of variable size in the structure provided 
each of the analyzed words will end with identical key. There are no 
obstacles for strings because they normally finish with ``end of line'' sign.
Owing to the reading of word keys and going through indexes (table look-at), 
primary, partial operations of PTrie method are very fast. If we carefully 
match $K$ with data type, PTrie will certainly serve as a really effective 
Priority Queue.\break
\subsection{Linear time algorithm finding the Minimal Spanning Tree (MST)}

\begin{myDefi}[MST~\cite{Cormen}]
Let $G = (V,E)$ be a connected, weighted, undirected graph. Any edges of graph 
$G = (V, E)$ have a weight function \mbox{$w : E \rightarrow R$}.  Spanning tree of 
$G$ is a subgraph $T$ which contains all of the graph's vertices.
The weight of a spanning tree $T$ is the sum of the weights of its edges:
\begin{displaymath}
w(T) = \sum_{E \in T}{w(E)}
\end{displaymath}
A minimum spanning tree of $G = (V, E)$ is acyclic subset $T \subseteq E$ that 
connects all of the vertices and whose total weight is minimized 
{\normalfont{[Figure~\ref{treeMST}]}}.
\end{myDefi}
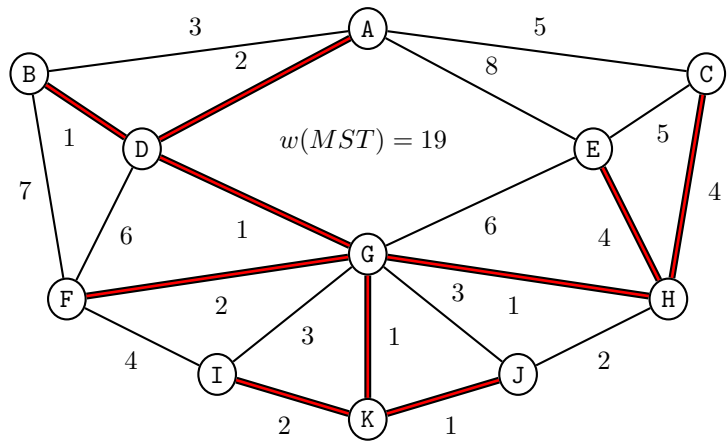
\begin{figure}[!hbp]
\caption{The Minimum Spinning Tree of $G = (V,E)$\label{treeMST}}
\begin{center}
\begin{pspicture}(0,0)(10,6)

\rput[bc](5, 4.1){$w(MST) = 19$ }

\ttfamily

\rput[bc](5, 5.6){\ovalnode{NodeA}{\textbf{A}}}
\rput[bc](0.5, 5){\ovalnode{NodeB}{\textbf{B}}}
\rput[bc](9.5, 5){\ovalnode{NodeC}{\textbf{C}}}
\rput[bc](2, 4){\ovalnode{NodeD}{\textbf{D}}}
\rput[bc](8, 4){\ovalnode{NodeE}{\textbf{E}}}
\rput[bc](5, 2.6){\ovalnode{NodeG}{\textbf{G}}}
\rput[bc](1, 2){\ovalnode{NodeF}{\textbf{F}}}
\rput[bc](9, 2){\ovalnode{NodeH}{\textbf{H}}}
\rput[bc](3, 1){\ovalnode{NodeI}{\textbf{I}}}
\rput[bc](7, 1){\ovalnode{NodeJ}{\textbf{J}}}
\rput[bc](5, 0.4){\ovalnode{NodeK}{\textbf{K}}}

\ncline{-}{NodeB}{NodeA}\Aput{$3$}
\ncline{-}{NodeA}{NodeC}\Aput{$5$}
\ncline[doubleline=true, doublecolor=red]{-}{NodeD}{NodeA}\Aput{$2$}
\ncline{-}{NodeA}{NodeE}\Aput{$8$}
\ncline[doubleline=true, doublecolor=red]{-}{NodeD}{NodeB}\Aput{$1$}
\ncline{-}{NodeF}{NodeB}\Aput{$7$}
\ncline[doubleline=true, doublecolor=red]{-}{NodeC}{NodeH}\Aput{$4$}
\ncline{-}{NodeC}{NodeE}\Aput{$5$}
\ncline[doubleline=true, doublecolor=red]{-}{NodeG}{NodeD}\Aput{$1$}
\ncline{-}{NodeD}{NodeF}\Aput{$6$}
\ncline[doubleline=true, doublecolor=red]{-}{NodeG}{NodeF}\Aput{$2$}
\ncline{-}{NodeI}{NodeF}\Aput{$4$}
\ncline[doubleline=true, doublecolor=red]{-}{NodeH}{NodeE}\Aput{$4$}
\ncline{-}{NodeH}{NodeJ}\Aput{$2$}
\ncline[doubleline=true, doublecolor=red]{-}{NodeH}{NodeG}\Aput{$1$}
\ncline{-}{NodeG}{NodeJ}\Aput{$3$}
\ncline[doubleline=true, doublecolor=red]{-}{NodeJ}{NodeK}\Aput{$1$}
\ncline[doubleline=true, doublecolor=red]{-}{NodeG}{NodeK}\Aput{$1$}
\ncline{-}{NodeG}{NodeI}\Aput{$3$}
\ncline{-}{NodeE}{NodeG}\Aput{$6$}
\ncline[doubleline=true, doublecolor=red]{-}{NodeK}{NodeI}\Aput{$2$}
\end{pspicture} 
\end{center}
\end{figure}
Let $A$ be a subset of $E$ that is included in some minimum spanning tree 
for $G = (V,E)$.  Jarnik-Prim's algorithm has the property 
that the edges in the set $A$ always form a single tree. The tree starts 
from an arbitrary source vertex $s$ and grows until the tree spans all the 
vertices in $V$. At each step, a light edge is added to the tree $A$ that 
connects $A$ to an isolated vertex of $G_A = (V, A)$.  
With the proof of Jarnik-Prim's algorithm follows that by using
this rule adds only edges that are safe for $A$; therefore, when the 
algorithm terminates, the edges in $A$ form a minimum spanning tree. 
This strategy is greedy since the tree is augmented at each step with an edge 
that contributes the minimum amount possible to the tree's weight.
The key to implementing Jarnik-Prim's algorithm efficiently is to make it easy 
to select a new edge to be added to the tree formed by the edges in $A$.
The performance of Jarnik-Prim's algorithm depends on how we implement 
the min-priority queue $Q$. If $Q$ is implemented as a binary min-heap, the 
total time for Jarnik-Prim's algorithm is 
$O(|V|log|V| + |E|log|V|) = O(|E|log|V|)$. 
\begin{lemi}
Using the priority queue based on multilevel prefix 
tree (PTrie) to implement the min-priority queue $Q$, the running time of 
Jarnik-Prim's algorithm improves to running worst time $O(|E|+|V|)$.
\end{lemi}
\proof
Algorithm crosses the graph adding one edge to subset $T$. All the edges are 
inserted to PTrie -- the structure working as the priority queue. In 
algorithm we use three operations of PTrie: insert, extract-min and 
decrease-key. Insert and decrease-key are characterized by 
$\Theta(\frac{M}{K} + K)$ time complexity, where $M$ is the length of key 
required to remember the weight of edge and $K$ is constant defined by 
programmers as the value of function $w(K) = (\frac{M}{K} + K)$ is 
minimized. Time complexity of extract-min is constant $\Theta(1)$.
If we use PTrie to set a successive arcs appending to subset $T$, by means 
of Jarnik-Prim's method, we gain time complexity which amounts to 
$\Theta(|V| + |E|*w(k))$. Let's assume that the size of word (word length) 
needed to remember the weight of arcs is constant for all arcs of the graph 
$G=(V,E)$, then function $w(k)$ is constant. We can calculate minimum 
coefficient of $min\{w(k)\}$ by matching suitably $K$ with $M$. Therefore 
time cost equals \mbox{$\Theta(|V| + |E|*min\{w_{const}(k)\}) = O(|V| + |E|)$}, 
where coefficient equals $min\{w(k) = (\frac{M{const}}{K}+K)\}$.
\subsection{Minimum-weight and minimal-vertex-amount path algorithm with 
            positive weights on directed graph in linear worst-case time (SSSP)}

\begin{myDefi}[SSSP~\cite{Cormen}]
In a Single-Source shortest-paths with positive weights problem, we are 
given a weighted, directed graph $G = (V, E)$, with weight function 
\mbox{$w : E \rightarrow R_+$} mapping edges to positive real-valued-weights. 
The weight of path\break \mbox{$p = <v_0, v_1, ..., v_k>$} is the sum of the 
weights of its constituent edges:
\begin{displaymath}
w(p) = \sum_{i=1}^{k} {(v_{i-1}, v_i)}
\end{displaymath}
\center{We define the shortest-path weight from u to v by}
\begin{displaymath}
\delta(u,v) = \left\{ \begin{array}{l} \verb@min@\{w(p) : u\stackrel{p}{\rightsquigarrow} v\}\\ 
\verb@path from u to v is not exist@\\ \end{array} \right.
\end{displaymath}
A shortest path from vertex $u$ to vertex $v$ is then defined as any existing 
path $p$ with weight $w(p) = \delta(u,v)$ \normalfont{[Figure~\ref{SSSPsourceS}]}.
\end{myDefi}
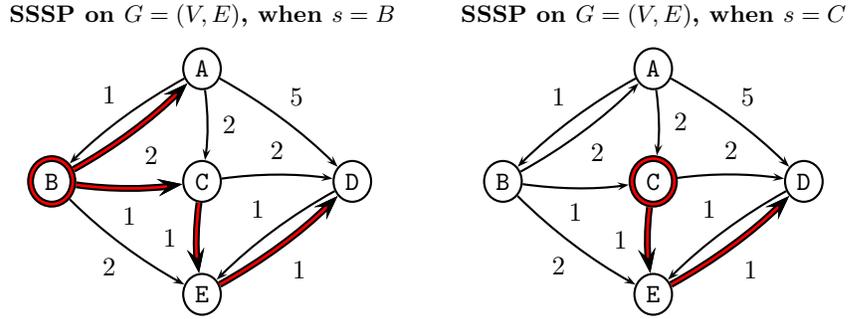
\begin{figure}[!hbp]
\caption{The Single-Source Shortest Path on directed graph $G = (V,E)$ for different source vertex $s$\label{SSSPsourceS}}
\begin{center}
\begin{pspicture}(0,0)(12,4.4)

\rput[bc](3, 4.2){ \small{\textbf{SSSP on $G = (V,E)$, when $s = B$}} }
\rput[bc](9, 4.2){ \small{\textbf{SSSP on $G = (V,E)$, when $s = C$}} }

\ttfamily

\rput[bc](3, 3.5){\ovalnode{NodeA}{\textbf{A}}}
\rput[bc](1, 2){\ovalnode[doubleline=true, doublecolor=red]{NodeB}{\textbf{B}}}
\rput[bc](3, 2){\ovalnode{NodeC}{\textbf{C}}}
\rput[bc](5, 2){\ovalnode{NodeD}{\textbf{D}}}
\rput[bc](3, 0.5){\ovalnode{NodeE}{\textbf{E}}}
\ncarc[doubleline=true, doublecolor=red]{<-}{NodeA}{NodeB}\Aput{$2$}
\ncarc{->}{NodeA}{NodeC}\Aput{$2$}
\ncarc{->}{NodeA}{NodeD}\Aput{$5$}
\ncarc{<-}{NodeB}{NodeA}\Aput{$1$}
\ncarc{->}{NodeC}{NodeD}\Aput{$2$}
\ncarc[doubleline=true, doublecolor=red]{<-}{NodeC}{NodeB}\Aput{$1$}
\ncarc[doubleline=true, doublecolor=red]{<-}{NodeD}{NodeE}\Aput{$1$}
\ncarc{<-}{NodeE}{NodeD}\Aput{$1$}
\ncarc{<-}{NodeE}{NodeB}\Aput{$2$}
\ncarc[doubleline=true, doublecolor=red]{<-}{NodeE}{NodeC}\Aput{$1$}

\rput[bc](9, 3.5){\ovalnode{NodeA}{\textbf{A}}}
\rput[bc](7, 2){\ovalnode{NodeB}{\textbf{B}}}
\rput[bc](9, 2){\ovalnode[doubleline=true, doublecolor=red]{NodeC}{\textbf{C}}}
\rput[bc](11, 2){\ovalnode{NodeD}{\textbf{D}}}
\rput[bc](9, 0.5){\ovalnode{NodeE}{\textbf{E}}}
\ncarc{<-}{NodeA}{NodeB}\Aput{$2$}
\ncarc{->}{NodeA}{NodeC}\Aput{$2$}
\ncarc{->}{NodeA}{NodeD}\Aput{$5$}
\ncarc{<-}{NodeB}{NodeA}\Aput{$1$}
\ncarc{->}{NodeC}{NodeD}\Aput{$2$}
\ncarc{<-}{NodeC}{NodeB}\Aput{$1$}
\ncarc[doubleline=true, doublecolor=red]{<-}{NodeD}{NodeE}\Aput{$1$}
\ncarc{<-}{NodeE}{NodeD}\Aput{$1$}
\ncarc{<-}{NodeE}{NodeB}\Aput{$2$}
\ncarc[doubleline=true, doublecolor=red]{<-}{NodeE}{NodeC}\Aput{$1$}

\end{pspicture} 
\end{center}
\end{figure}
Dijkstra's algorithm maintains a set $S$ of vertices whose final 
shortest-path weights from the source $s$ have already been determined. 
The algorithm repeatedly selects the vertex $u \in V - S$ with the 
minimum shortest-path estimate, adds $u$ to $S$, and relaxes all edges 
leaving $u$. The running time of Dijkstra's algorithm depends on how 
the min-priority queue is implemented. The performance of Dijkstra's 
algorithm depends on how we implement the min-priority queue $Q$. 
If $Q$ is implemented as a binary min-heap, the total time for 
Dijkstra's algorithm is $O((|V|+|E|)log|V|) = O(|E|log|V|)$.\break

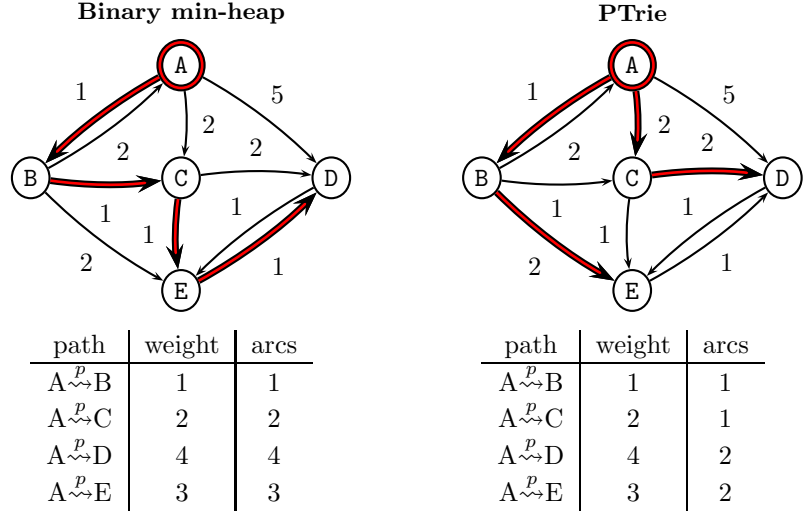
\begin{figure}[!]
\caption{The difference of Dijkstry's algorithm between the use of basic priority queue and PTrie\label{DifferenceBvsPTrie}}
\begin{center}
\begin{pspicture*}(0,0)(12,4.4)

\rput[bc](3, 4.2){ \small{\textbf{Binary min-heap}} }
\rput[bc](9, 4.2){ \small{\textbf{PTrie}} }

\ttfamily

\rput[bc](3, 3.5){\ovalnode[doubleline=true, doublecolor=red]{NodeA}{\textbf{A}}}
\rput[bc](1, 2){\ovalnode{NodeB}{\textbf{B}}}
\rput[bc](3, 2){\ovalnode{NodeC}{\textbf{C}}}
\rput[bc](5, 2){\ovalnode{NodeD}{\textbf{D}}}
\rput[bc](3, 0.5){\ovalnode{NodeE}{\textbf{E}}}
\ncarc{<-}{NodeA}{NodeB}\Aput{$2$}
\ncarc{->}{NodeA}{NodeC}\Aput{$2$}
\ncarc{->}{NodeA}{NodeD}\Aput{$5$}
\ncarc[doubleline=true, doublecolor=red]{<-}{NodeB}{NodeA}\Aput{$1$}
\ncarc{->}{NodeC}{NodeD}\Aput{$2$}
\ncarc[doubleline=true, doublecolor=red]{<-}{NodeC}{NodeB}\Aput{$1$}
\ncarc[doubleline=true, doublecolor=red]{<-}{NodeD}{NodeE}\Aput{$1$}
\ncarc{<-}{NodeE}{NodeD}\Aput{$1$}
\ncarc{<-}{NodeE}{NodeB}\Aput{$2$}
\ncarc[doubleline=true, doublecolor=red]{<-}{NodeE}{NodeC}\Aput{$1$}

\rput[bc](9, 3.5){\ovalnode[doubleline=true, doublecolor=red]{NodeA}{\textbf{A}}}
\rput[bc](7, 2){\ovalnode{NodeB}{\textbf{B}}}
\rput[bc](9, 2){\ovalnode{NodeC}{\textbf{C}}}
\rput[bc](11, 2){\ovalnode{NodeD}{\textbf{D}}}
\rput[bc](9, 0.5){\ovalnode{NodeE}{\textbf{E}}}
\ncarc{<-}{NodeA}{NodeB}\Aput{$2$}
\ncarc[doubleline=true, doublecolor=red]{->}{NodeA}{NodeC}\Aput{$2$}
\ncarc{->}{NodeA}{NodeD}\Aput{$5$}
\ncarc[doubleline=true, doublecolor=red]{<-}{NodeB}{NodeA}\Aput{$1$}
\ncarc[doubleline=true, doublecolor=red]{->}{NodeC}{NodeD}\Aput{$2$}
\ncarc{<-}{NodeC}{NodeB}\Aput{$1$}
\ncarc{<-}{NodeD}{NodeE}\Aput{$1$}
\ncarc{<-}{NodeE}{NodeD}\Aput{$1$}
\ncarc[doubleline=true, doublecolor=red]{<-}{NodeE}{NodeB}\Aput{$2$}
\ncarc{<-}{NodeE}{NodeC}\Aput{$1$}
\end{pspicture*} 
\begin{tabular}{ p{4cm} @{} p{2cm} @{} p{4cm}}
 \begin{tabular}{c|c|c}
   path & weight & arcs\\
   \hline
   A$\stackrel{p}{\rightsquigarrow}$B & 1 & 1\\
   A$\stackrel{p}{\rightsquigarrow}$C & 2 & 2\\
   A$\stackrel{p}{\rightsquigarrow}$D & 4 & 4\\
   A$\stackrel{p}{\rightsquigarrow}$E & 3 & 3
 \end{tabular}
 &  
 & 
 \begin{tabular}{c|c|c}
   path & weight & arcs\\
   \hline
   A$\stackrel{p}{\rightsquigarrow}$B & 1 & 1\\
   A$\stackrel{p}{\rightsquigarrow}$C & 2 & 1\\
   A$\stackrel{p}{\rightsquigarrow}$D & 4 & 2\\
   A$\stackrel{p}{\rightsquigarrow}$E & 3 & 2
 \end{tabular}
 \\

\end{tabular}

\end{center}
\end{figure}

The quest for linear worst-case time Single-Source Shortest Path Algorithm 
on arbitrary directed graphs with positive arc weights is on ongoing hot 
research topic. Algorithm which I present not only finds minimum-weight 
path (shortest), but also makes the path walk through as few vertices as 
possible. I propose implementation of Dijkstra's algorithm which uses 
priority queue $Q$ based on multilevel prefix tree (PTrie) 
[Figure~\ref{DifferenceBvsPTrie}]. PTrie is a stable structure~\cite{Planeta06}. 
Thanks to this algorithm it not only builds the Shortest Path of 
minimum-weight considering the arc weights, but also considering to the 
number of vertices.
\begin{lemi}
Dijkstra's algorithm where PTrie is used by priority queue request\\ 
\mbox{$O(|V| + |E|)$} time.
\end{lemi}
\proof
Dijkstra's algorithm makes use of tree operations of PTrie: 
insert, extract-min and remove of  $\Theta(|V| * |E|w(k))$ time cost. 
Because the length of word (size) necessary to remember the weight of arcs 
is constant for all arcs of graph $G=(V,E)$, function $w(k)$ is constant. 
Function $w(k)$ is a constant coefficient which equals 
\mbox{$min\{w(k) = (\frac{Mconst}{K} + K)\}$}. Which means that time cost of 
particular operations executed by PTrie in case of SSSP problem equals $O(1)$. 
Therefore Dijkstra's algorithm where PTrie is used by priority queue needs 
$O(|V| + |E|)$ time.
\subsection{Single-Destination Shortest-Paths problem (SDSP) needs linear time}
The reverse SSSP problem is commonly used in practice. Find a shortest path to 
a given destination source vertex $s$ from each vertex $v$ (SDSP). 
By reversing the direction of each edge in the graph $G=(V,E)$, we can 
reduce this problem to a single-source problem
{\normalfont{[Figure~\ref{DifferenceSSSPvsSDSP}]}}. 
Algorithm build SDSP tree $T$ (subset of graph) of shortest paths to source 
vertex from each vertex, whose leaves are all vertices $v \in V$ -- without 
the initial source vertex $s \in V$, which is the root of $T$. All paths 
lead from each arbitrary vertex $v \in V$ to source vertex $s$ for vertices 
accessible from a given destination source vertex $s$ .
\begin{figure}[!]
\caption{Difference between SSSP and SDSP problems\label{DifferenceSSSPvsSDSP}}
\begin{center}
\begin{pspicture*}(0,0)(12,5.4)

\rput[bc](2.5, 5.2){ \small{\textbf{Single-Source Shortest Path}} }
\rput[bc](9.5, 5.2){ \small{\textbf{Single-Destination Shortest-Path}} }

\ttfamily

\rput[bc](2.5, 4.5){\ovalnode[doubleline=true, doublecolor=red]{NodeA}{\textbf{A}}}
\rput[bc](1, 3.5){\ovalnode{NodeB}{\textbf{B}}}
\rput[bc](4, 3.5){\ovalnode{NodeC}{\textbf{C}}}
\rput[bc](2.5, 2.5){\ovalnode{NodeD}{\textbf{D}}}
\rput[bc](0.5, 1.5){\ovalnode{NodeE}{\textbf{E}}}
\rput[bc](4.5, 1.5){\ovalnode{NodeF}{\textbf{F}}}
\rput[bc](2.5, 0.5){\ovalnode{NodeG}{\textbf{G}}}

\ncline[doubleline=true, doublecolor=red]{<-}{NodeB}{NodeA}\Aput{$1$}
\ncline[doubleline=true, doublecolor=red]{->}{NodeA}{NodeC}\Aput{$4$}
\ncline[doubleline=true, doublecolor=red]{->}{NodeA}{NodeD}\Aput{$3$}
\ncline{-}{NodeC}{NodeD}\Aput{$6$}
\ncline{-}{NodeC}{NodeF}\Aput{$1$}
\ncline{-}{NodeD}{NodeB}\Aput{$4$}
\ncline{-}{NodeD}{NodeE}\Aput{$2$}
\ncline[doubleline=true, doublecolor=red]{<-}{NodeE}{NodeB}\Aput{$2$}
\ncline[doubleline=true, doublecolor=red]{<-}{NodeF}{NodeD}\Aput{$1$}
\ncline[doubleline=true, doublecolor=red]{->}{NodeF}{NodeG}\Aput{$1$}
\ncline{-}{NodeG}{NodeE}\Aput{$5$}

\rput[bc](9.5, 4.5){\ovalnode[doubleline=true, doublecolor=red]{NodeA}{\textbf{A}}}
\rput[bc](8, 3.5){\ovalnode{NodeB}{\textbf{B}}}
\rput[bc](11, 3.5){\ovalnode{NodeC}{\textbf{C}}}
\rput[bc](9.5, 2.5){\ovalnode{NodeD}{\textbf{D}}}
\rput[bc](7.5, 1.5){\ovalnode{NodeE}{\textbf{E}}}
\rput[bc](11.5, 1.5){\ovalnode{NodeF}{\textbf{F}}}
\rput[bc](9.5, 0.5){\ovalnode{NodeG}{\textbf{G}}}

\ncline[doubleline=true, doublecolor=red]{->}{NodeB}{NodeA}\Aput{$1$}
\ncline[doubleline=true, doublecolor=red]{<-}{NodeA}{NodeC}\Aput{$4$}
\ncline[doubleline=true, doublecolor=red]{<-}{NodeA}{NodeD}\Aput{$3$}
\ncline{-}{NodeC}{NodeD}\Aput{$6$}
\ncline{-}{NodeC}{NodeF}\Aput{$1$}
\ncline{-}{NodeD}{NodeB}\Aput{$4$}
\ncline{-}{NodeD}{NodeE}\Aput{$2$}
\ncline[doubleline=true, doublecolor=red]{->}{NodeE}{NodeB}\Aput{$2$}
\ncline[doubleline=true, doublecolor=red]{->}{NodeF}{NodeD}\Aput{$1$}
\ncline[doubleline=true, doublecolor=red]{<-}{NodeF}{NodeG}\Aput{$1$}
\ncline{-}{NodeG}{NodeE}\Aput{$5$}

\end{pspicture*} 
\end{center}
\end{figure}
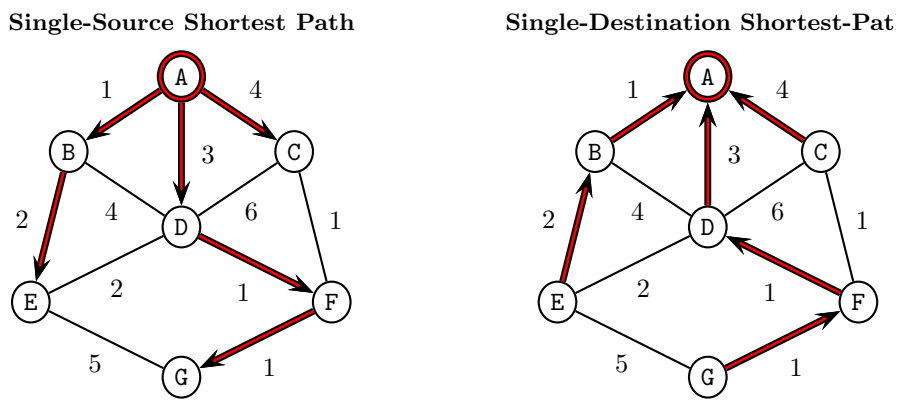
\begin{myDefi}[SDSP]
We are given a weighted, directed graph $G = (V,E)$, represented by adjacent 
list, with weight function $w : E\rightarrow R_{+}$ mapping edges to 
positive real-valued-weights. The weight of path 
$p = <v_0, v_1, \ldots, v_k>$, is the sum of the weights of its constituent 
edges. For graph $G = (V,E)$ exist destination source vertex $s \in V$; 
for all vertices $v \in V$ it is necessary to find the shortest path 
with $v$ to $s$.
\end{myDefi}
Similarly to other algorithms I use the property that Shortest Paths 
algorithms typically rely on the property that the shortest path between 
two vertices contains other shortest paths within it. But I assume double 
criterion to build the shortest path. I build the shortest path relative to 
the weight of arcs and then relative to the amount of vertices which contain 
the shortest (minimum-weight and minimum-vertices) path. That is possible 
thanks to the stability of PTrie implementation~\cite{Planeta06}. That 
solution is suitable, because it may happen that there exist many shortest 
paths related to the weight of arcs. In these circumstances, 
minimal-weight path with minimum amount of arcs becomes the Shortest Path.
\subsubsection{Structure of vertex and arc}

\begin{figure}[!]
\caption{SDSP tree\label{SDSPtree}}
\begin{center}
\begin{pspicture*}(0,0)(6.8, 3.5)
\pcline{<-}(3.05, 3)(6.6, 0.5)
\aput*{:U}{Shortest\space path $p(v_k, v_s)$}

\ttfamily
\rput[bc](4.7, 0.5){$\ldots$}
\rput[bc](2,3.2){\rnode{NIL}{NIL}}

\rput[bc](1.2,2.6){\rnode{BACK}{$\stackrel{v_{1}.back}{\rightsquigarrow}$}}
\rput[bl](0.1,2.4){\rnode{b1}{}}
\rput[bl](2.1,2.4){\rnode{b2}{}}
\ncline{-}{b1}{b2}
\rput[bl](2.2,2.1){\rnode{b3}{}}
\ncline{->}{b2}{b3}

\rput[bc](3, 2.5){\ovalnode[doubleline=true, doublecolor=red]{NodeS}{$v_s$}}
\rput[bc](1.5, 1.5){\ovalnode{NodeV1}{$v_1$}}
\rput[bc](4.5, 1.5){\ovalnode{NodeV2}{$v_2$}}
\rput[bc](0.5, 0.5){\ovalnode{NodeV3}{$v_3$}}
\rput[bc](1.5, 0.5){\ovalnode{NodeV4}{$v_4$}}
\rput[bc](2.5, 0.5){\ovalnode{NodeV5}{$v_5$}}
\rput[bc](3.5, 0.5){\ovalnode{NodeV6}{$v_6$}}
\rput[bc](5.9, 0.5){\ovalnode{NodeVK}{$v_k$}}
\ncline{->}{NodeV1}{NodeS}
\ncline{->}{NodeV2}{NodeS}
\ncline{->}{NodeV3}{NodeV1}
\ncline{->}{NodeV4}{NodeV1}
\ncline{->}{NodeV5}{NodeV1}
\ncline{->}{NodeV6}{NodeV2}
\ncline{->}{NodeVK}{NodeV2}
\ncline{->}{NodeS}{NIL}
\end{pspicture*}
\end{center}
\end{figure}
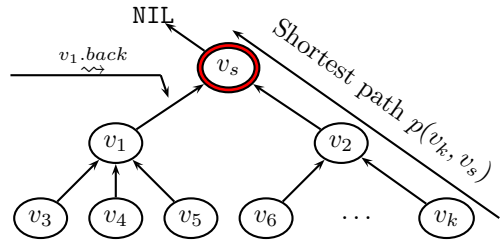

Structure of vertex contain type `$data$' which storage label of vertex. Because graph
\begin{tabular}{ l @{} p{0.1cm} @{} p{9.2cm}}
 \begin{tabular}{|l|}
 \hline
 \verb@Vertex{@\\
 \verb@data;@\\
 \verb@Neighbors *list;@\\
 \verb@Neighbors *back;@\\
 \verb@};@\\
 \hline
 \end{tabular}
 &
 &
 \begin{tabular}{p{9.2cm}}
 $G$ is represented by adjacent list, each vertex has the  
 list of pointers to the neighbors. The order on the list is random. The 
 structure of vertex has a helpful variable `$back$' (used by any graph 
 algorithms), which indicates one of arcs locate on the adjacent list of 
 the structure. Each vertex $v \in V$ has a
 \end{tabular}
\end{tabular}
\\
link `$back$' to the neighbor (vertex from the adjacent list), in that 
moment considered the successor on the shortest path, or `$back$' 
equals NIL. For this reason we can for example, differentiate the 
vertices added to SDSP tree $T$ {\normalfont{[Figure~\ref{SDSPtree}]}} 
from those ones which hasn't been analyzed yet.
\\
\\
The structure of arc serves to insert information about arc to PTrie.
Therefore PTrie\\
\begin{tabular}{l @{} p{0.1cm} @{} p{9.2cm}}
 \begin{tabular}{|l|}
 \hline
 \verb@Arc{@\\
 \verb@weight;@\\
 \verb@pathWeight;@\\
 \verb@Vertex *tail;@\\
 \verb@Vertex *head;@\verb@    @\\
 \verb@};@\\
 \hline
 \end{tabular}
&
&
 \begin{tabular}{p{9.2cm}}
will be able to store not only integer but detail 
information about arbitrary arc too. The Arc structure stores information 
about the weight of arc and path, and two pointers; to vertex which is the 
tail of the arc and to vertex which is the head of the arc. The `$pathWeight$' 
contains the sum of optimal arcs, which follow from the source vertex to 
currently
 \end{tabular}
\end{tabular}
\\
analyzed vertex. PTrie uses this variable to determine the order.
\\

Such graph implementation allows considerable adaptability. It's enough to 
know the address of vertex (pointer) 
to get to know the 
shortest path to a given destination of the source vertex. And by the way 
meet all vertices which are located on this path.\break
\subsubsection{Algorithm}
The algorithm starts with the source vertex $s \in V$ and inserts the 
adjacent list of `$s$' to PTrie. Then with the help of `$minimum$' and 
`$remove$' operations `$extracts-remove-min$' of arcs from PTrie.
We move on to the arc leadings to vertex. If vertex has not been attached 
to SDSP tree yet (the value of back is equal $NIL$) the algorithm will 
attach the vertex to SDSP tree. By setting the pointer `$back$' on the tail 
(vertex) of the arc which leads to the current vertex. Next, all arcs of 
the analyzed vertex increased by the weight of the path, which leads to the 
current vertex, are inserted to PTrie. Again we choose the smallest arc 
from PTrie$\ldots$ The algorithm ends its work when PTrie is empty. It means 
that all arcs accessible from the  source vertex were browsed. Visited 
vertices have set arc `$back$' is such a way that the path which the arcs 
`$back$' built is not only the minimum-weight path (the amount of arc 
weights is the smallest) but also the path walks through as few arcs as 
possible.\break
\begin{center}
Pseudo-code of algorithm compute SDSP problem\\
\begin{tabular}{|c|l|}
 \hline
  (1) & \verb@SDSP(G, s) begin@\\
  (2) & \verb@   PTrie.insert(s.@$\stackrel{------\rightarrow}{neighbors}$\verb@)@\\
  (3) & \verb@   while(PTrie is not empty) begin@\\
  (4) & \verb@      arc = PTrie.minimum()@\\
  (5) & \verb@      PTrie.remove(arc)@\\
  (6) & \verb@      if(arc.head.back is empty and isn't s) begin@\\
  (7) & \verb@         arc.head.back = reverse(arc)@\\
  (8) & \verb@         PTrie.insert(arc.head.@$\stackrel{------\rightarrow}{neighbors}$\verb@ + arc.pathWeight)@\\
  (9) & \verb@      end@\\
 (10) & \verb@   end@\\  
      & \\
 (11) & \verb@end@\\        
 \hline
\end{tabular}
\begin{enumerate}
	\item The algorithm begins to build the SDSP tree from the arbitrary source 
	vertex `$s$'. SDSP tree consists of all vertices accessible from any source 
	vertex.
	\item Insert the adjacent list to PTrie.
	\item The algorithm will check the paths stored in PTrie as long as they 
	exist.
	\item I take and remember the path of the smallest weight from PTrie and 
	the last arc of this path. The variable `$weightPath$' defines the weight 
	of the whole path. The variable `$weight$' defines the weight of the last 
	arc, where the last arc is represented by variables `$tail$' and `$head$'. 
	The path leads from the source vertex to the vertex indicate by `$head$'
	\item Remove the path of the smallest weight from PTrie.
	\item If the vertex which the arc leads to has not been added to SDSP 
	tree yet and it is not the source vertex$\ldots$
	\item Ascribe the reverse of analyzed arc to the supportive arc `$back$'.\\
	\verb@arc:@ $A\rightarrow B$\\
  \verb@reverse(arc):@ $B\rightarrow A$\\
  \item Insert the arc of analyzed vertex to PTrie adding the weight of the 
  path which brought us to the analyzed vertex.
  \item If the vertex has already been added to SDSP tree or its is a 
  source vertex, it is not analyzed any more.
  \item The algorithm finished checking all arcs/vertices which were 
  accessible from the source vertex.
  \item When the algorithm finishes its work an vertices accessible from 
  the source vertex by the supportive arcs `$back$' build SDSP tree, whose 
  root and vertex constitute the source vertex, to which lead all the paths 
  based on the arcs `$back$'.
\end{enumerate}
\end{center}
\subsubsection{Analysis of the algorithm work}
I will analyze the algorithm work step by step; How and in what order arcs 
are inserted to PTrie? What is the sequence of vertex attachment to the tree 
containing the solution of SDSP problem? Step by step description of the 
algorithm computing SDSP problem at work
{\normalfont{[Figures~\ref{SDSPtree1},\ref{SDSPtree2},\ref{SDSPtree3},\ref{SDSPtree4},%
\ref{SDSPtree5},\ref{SDSPtree6},\ref{SDSPtree7},\ref{SDSPtree8},\ref{SDSPtree9},%
\ref{SDSPtree10},\ref{SDSPtree11}]}}. 
\begin{figure}[!]
\caption{Algorithm compute SDSP of a work, step I\label{SDSPtree1}}
\begin{center}
\begin{tabular}{ p{6.3cm} @{} c}
\begin{tabular}{l}
\begin{pspicture*}(-0.3,-0.2)(6, 5)

\rput[bc](-0.15, 2.5){ \rotateleft{$Step$ $I$} }

\ttfamily
\rput[bc](1.5, 4.5){\ovalnode[doubleline=true,doublecolor=red, fillstyle=solid,fillcolor=yellow]{NodeA}{\textbf{A}}}
\rput[bc](4, 4){\ovalnode{NodeB}{\textbf{B}}}
\rput[bc](0.5, 3){\ovalnode{NodeC}{\textbf{C}}}
\rput[bc](2.5, 2){\ovalnode{NodeD}{\textbf{D}}}
\rput[bc](5.5, 2.5){\ovalnode{NodeE}{\textbf{E}}}
\rput[bc](1, 0.5){\ovalnode{NodeF}{\textbf{F}}}
\rput[bc](5, 0.5){\ovalnode{NodeG}{\textbf{G}}}

\ncarc{->}{NodeA}{NodeB}\Aput{$3$}
\ncarc{->}{NodeA}{NodeD}\Aput{$1$}
\ncarc{->}{NodeA}{NodeC}\Aput{$5$}

\ncarc{->}{NodeB}{NodeA}\Aput{$1$}
\ncarc{->}{NodeB}{NodeE}\Aput{$4$}

\ncarc{->}{NodeC}{NodeA}\Aput{$1$}
\ncarc{->}{NodeC}{NodeF}\Aput{$1$}

\ncarc{->}{NodeD}{NodeF}\Aput{$2$}
\ncarc{->}{NodeD}{NodeB}\Aput{$1$}
\ncarc{->}{NodeD}{NodeE}\Aput{$7$}
\ncarc{<-}{NodeB}{NodeD}\Aput{$3$}

\ncarc{->}{NodeE}{NodeD}\Aput{$2$}
\ncarc{->}{NodeE}{NodeG}\Aput{$3$}

\ncarc{->}{NodeG}{NodeE}\Aput{$0$}
\ncarc{->}{NodeG}{NodeF}\Aput{$2$}

\ncarc{->}{NodeF}{NodeC}\Aput{$1$}
\ncarc{->}{NodeF}{NodeG}\Aput{$1$}
\end{pspicture*}
\end{tabular}
&
\begin{tabular}{c}
 Inside the PTrie:\\
 \\
 \begin{tabular}{|c|c|}
  \hline
  weight & path\\
  \hline\hline
  1 & A$\stackrel{p}{\rightsquigarrow}$D\\ \hline
  3 & A$\stackrel{p}{\rightsquigarrow}$B\\ \hline
  5 & A$\stackrel{p}{\rightsquigarrow}$C\\ \hline
 \end{tabular}
 \\
 \\
 \begin{tabular}{l}
  extract-min: A$\stackrel{p}{\rightsquigarrow}$D\\
  path weight: $w(p) = 1$\\
  path reject: \verb@FALSE@\\
 \end{tabular}\\
\end{tabular}
\\
\end{tabular}
\end{center}
\end{figure}
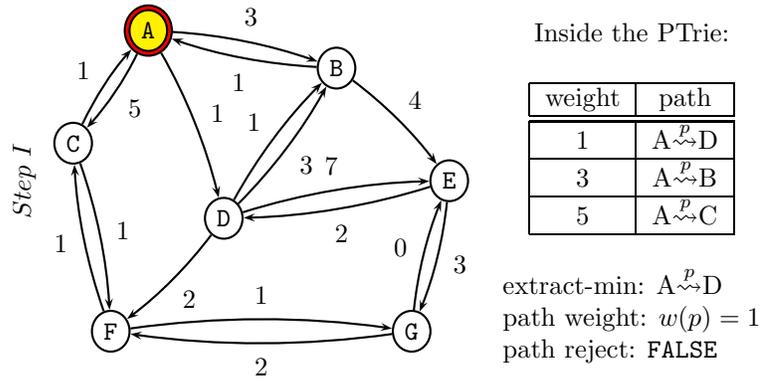
\begin{figure}[!]
\caption{Algorithm compute SDSP of a work, step II\label{SDSPtree2}}
\begin{center}
\begin{tabular}{ p{6.3cm} @{} c}
\begin{tabular}{l}
\begin{pspicture*}(-0.3,-0.2)(6, 5)

\rput[bc](-0.15, 2.5){ \rotateleft{$Step$ $II$} }

\ttfamily
\rput[bc](1.5, 4.5){\ovalnode[doubleline=true,doublecolor=red, fillstyle=solid,fillcolor=yellow]{NodeA}{\textbf{A}}}
\rput[bc](4, 4){\ovalnode{NodeB}{\textbf{B}}}
\rput[bc](0.5, 3){\ovalnode{NodeC}{\textbf{C}}}
\rput[bc](2.5, 2){\ovalnode[fillstyle=solid,fillcolor=yellow]{NodeD}{\textbf{D}}}
\rput[bc](5.5, 2.5){\ovalnode{NodeE}{\textbf{E}}}
\rput[bc](1, 0.5){\ovalnode{NodeF}{\textbf{F}}}
\rput[bc](5, 0.5){\ovalnode{NodeG}{\textbf{G}}}

\ncarc{->}{NodeA}{NodeB}\Aput{$3$}
\ncarc[doubleline=true,doublecolor=red]{->}{NodeA}{NodeD}\Aput{$1$}
\ncarc{->}{NodeA}{NodeC}\Aput{$5$}

\ncarc{->}{NodeB}{NodeA}\Aput{$1$}
\ncarc{->}{NodeB}{NodeE}\Aput{$4$}

\ncarc{->}{NodeC}{NodeA}\Aput{$1$}
\ncarc{->}{NodeC}{NodeF}\Aput{$1$}

\ncarc{->}{NodeD}{NodeF}\Aput{$2$}
\ncarc{->}{NodeD}{NodeB}\Aput{$1$}
\ncarc{->}{NodeD}{NodeE}\Aput{$7$}
\ncarc{<-}{NodeB}{NodeD}\Aput{$3$}

\ncarc{->}{NodeE}{NodeD}\Aput{$2$}
\ncarc{->}{NodeE}{NodeG}\Aput{$3$}

\ncarc{->}{NodeG}{NodeE}\Aput{$0$}
\ncarc{->}{NodeG}{NodeF}\Aput{$2$}

\ncarc{->}{NodeF}{NodeC}\Aput{$1$}
\ncarc{->}{NodeF}{NodeG}\Aput{$1$}
\end{pspicture*}
\end{tabular}
&
\begin{tabular}{c}
 Inside the PTrie:\\
 \\
 \begin{tabular}{|c|c|}
  \hline
  weight & path\\
  \hline\hline
  2 & D$\stackrel{p}{\rightsquigarrow}$B\\ \hline
  3 & A$\stackrel{p}{\rightsquigarrow}$B\\ \hline
  3 & D$\stackrel{p}{\rightsquigarrow}$F\\ \hline
  4 & D$\stackrel{p}{\rightsquigarrow}$B\\ \hline
  5 & A$\stackrel{p}{\rightsquigarrow}$C\\ \hline
  8 & D$\stackrel{p}{\rightsquigarrow}$E\\ \hline
 \end{tabular}
 \\
 \\
 \begin{tabular}{l}
  extract-min: D$\stackrel{p}{\rightsquigarrow}$B\\
  path weight: $w(p) = 2$\\
  path reject: \verb@FALSE@\\
 \end{tabular}\\
\end{tabular}
\\
\end{tabular}
\end{center}
\end{figure}
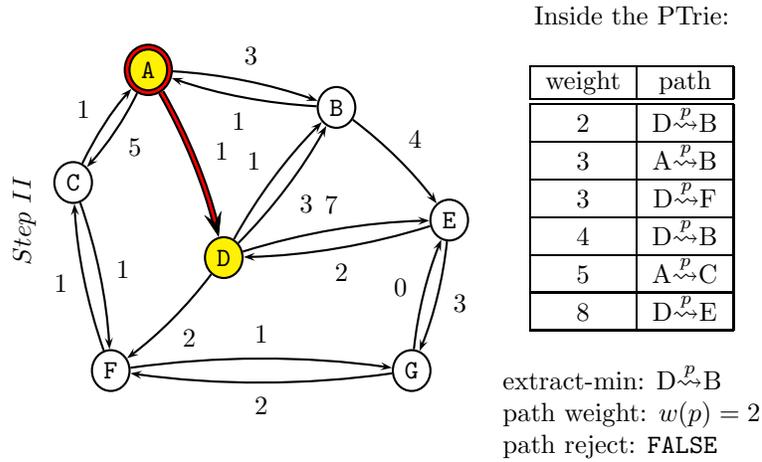
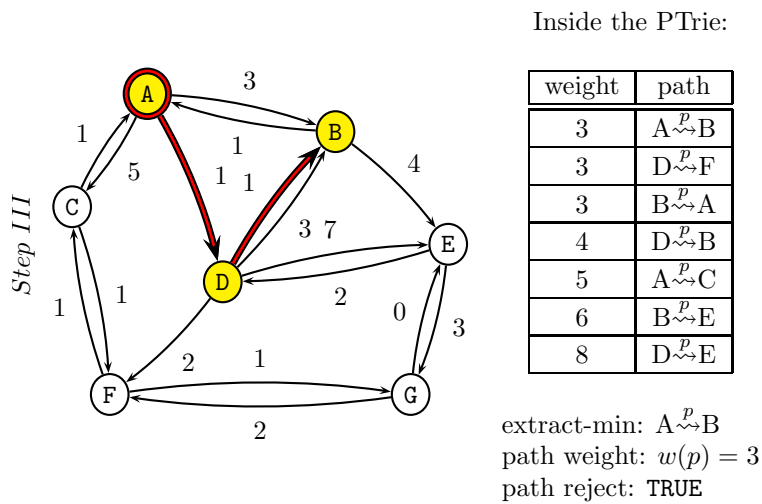
\begin{figure}[!]
\caption{Algorithm compute SDSP of a work, step III\label{SDSPtree3}}
\begin{center}
\begin{tabular}{ p{6.3cm} @{} c}
\begin{tabular}{l}
\begin{pspicture*}(-0.3,-0.2)(6, 5)

\rput[bc](-0.15, 2.5){ \rotateleft{$Step$ $III$} }

\ttfamily
\rput[bc](1.5, 4.5){\ovalnode[doubleline=true,doublecolor=red, fillstyle=solid,fillcolor=yellow]{NodeA}{\textbf{A}}}
\rput[bc](4, 4){\ovalnode[fillstyle=solid,fillcolor=yellow]{NodeB}{\textbf{B}}}
\rput[bc](0.5, 3){\ovalnode{NodeC}{\textbf{C}}}
\rput[bc](2.5, 2){\ovalnode[fillstyle=solid,fillcolor=yellow]{NodeD}{\textbf{D}}}
\rput[bc](5.5, 2.5){\ovalnode{NodeE}{\textbf{E}}}
\rput[bc](1, 0.5){\ovalnode{NodeF}{\textbf{F}}}
\rput[bc](5, 0.5){\ovalnode{NodeG}{\textbf{G}}}

\ncarc{->}{NodeA}{NodeB}\Aput{$3$}
\ncarc[doubleline=true,doublecolor=red]{->}{NodeA}{NodeD}\Aput{$1$}
\ncarc{->}{NodeA}{NodeC}\Aput{$5$}

\ncarc{->}{NodeB}{NodeA}\Aput{$1$}
\ncarc{->}{NodeB}{NodeE}\Aput{$4$}

\ncarc{->}{NodeC}{NodeA}\Aput{$1$}
\ncarc{->}{NodeC}{NodeF}\Aput{$1$}

\ncarc{->}{NodeD}{NodeF}\Aput{$2$}
\ncarc[doubleline=true,doublecolor=red]{->}{NodeD}{NodeB}\Aput{$1$}
\ncarc{->}{NodeD}{NodeE}\Aput{$7$}
\ncarc{<-}{NodeB}{NodeD}\Aput{$3$}

\ncarc{->}{NodeE}{NodeD}\Aput{$2$}
\ncarc{->}{NodeE}{NodeG}\Aput{$3$}

\ncarc{->}{NodeG}{NodeE}\Aput{$0$}
\ncarc{->}{NodeG}{NodeF}\Aput{$2$}

\ncarc{->}{NodeF}{NodeC}\Aput{$1$}
\ncarc{->}{NodeF}{NodeG}\Aput{$1$}
\end{pspicture*}
\end{tabular}
&
\begin{tabular}{c}
 Inside the PTrie:\\
 \\
 \begin{tabular}{|c|c|}
  \hline
  weight & path\\
  \hline\hline
  3 & A$\stackrel{p}{\rightsquigarrow}$B\\ \hline
  3 & D$\stackrel{p}{\rightsquigarrow}$F\\ \hline
  3 & B$\stackrel{p}{\rightsquigarrow}$A\\ \hline
  4 & D$\stackrel{p}{\rightsquigarrow}$B\\ \hline
  5 & A$\stackrel{p}{\rightsquigarrow}$C\\ \hline
  6 & B$\stackrel{p}{\rightsquigarrow}$E\\ \hline
  8 & D$\stackrel{p}{\rightsquigarrow}$E\\ \hline
 \end{tabular}
 \\
 \\
 \begin{tabular}{l}
  extract-min: A$\stackrel{p}{\rightsquigarrow}$B\\
  path weight: $w(p) = 3$\\
  path reject: \verb@TRUE@\\
 \end{tabular}\\
\end{tabular}
\\
\end{tabular}
\end{center}
\end{figure}
\begin{figure}[!]
\caption{Algorithm compute SDSP of a work, step IV\label{SDSPtree4}}
\begin{center}
\begin{tabular}{ p{6.3cm} @{} c}
\begin{tabular}{l}
\begin{pspicture*}(-0.3,-0.2)(6, 5)

\rput[bc](-0.15, 2.5){ \rotateleft{$Step$ $IV$} }

\ttfamily
\rput[bc](1.5, 4.5){\ovalnode[doubleline=true,doublecolor=red, fillstyle=solid,fillcolor=yellow]{NodeA}{\textbf{A}}}
\rput[bc](4, 4){\ovalnode[fillstyle=solid,fillcolor=yellow]{NodeB}{\textbf{B}}}
\rput[bc](0.5, 3){\ovalnode{NodeC}{\textbf{C}}}
\rput[bc](2.5, 2){\ovalnode[fillstyle=solid,fillcolor=yellow]{NodeD}{\textbf{D}}}
\rput[bc](5.5, 2.5){\ovalnode{NodeE}{\textbf{E}}}
\rput[bc](1, 0.5){\ovalnode{NodeF}{\textbf{F}}}
\rput[bc](5, 0.5){\ovalnode{NodeG}{\textbf{G}}}

\ncarc{->}{NodeA}{NodeB}\Aput{$3$}
\ncarc[doubleline=true,doublecolor=red]{->}{NodeA}{NodeD}\Aput{$1$}
\ncarc{->}{NodeA}{NodeC}\Aput{$5$}

\ncarc{->}{NodeB}{NodeA}\Aput{$1$}
\ncarc{->}{NodeB}{NodeE}\Aput{$4$}

\ncarc{->}{NodeC}{NodeA}\Aput{$1$}
\ncarc{->}{NodeC}{NodeF}\Aput{$1$}

\ncarc{->}{NodeD}{NodeF}\Aput{$2$}
\ncarc[doubleline=true,doublecolor=red]{->}{NodeD}{NodeB}\Aput{$1$}
\ncarc{->}{NodeD}{NodeE}\Aput{$7$}
\ncarc{<-}{NodeB}{NodeD}\Aput{$3$}

\ncarc{->}{NodeE}{NodeD}\Aput{$2$}
\ncarc{->}{NodeE}{NodeG}\Aput{$3$}

\ncarc{->}{NodeG}{NodeE}\Aput{$0$}
\ncarc{->}{NodeG}{NodeF}\Aput{$2$}

\ncarc{->}{NodeF}{NodeC}\Aput{$1$}
\ncarc{->}{NodeF}{NodeG}\Aput{$1$}
\end{pspicture*}
\end{tabular}
&
\begin{tabular}{c}
 Inside the PTrie:\\
 \\
 \begin{tabular}{|c|c|}
  \hline
  weight & path\\
  \hline\hline
  3 & D$\stackrel{p}{\rightsquigarrow}$F\\ \hline
  3 & B$\stackrel{p}{\rightsquigarrow}$A\\ \hline
  4 & D$\stackrel{p}{\rightsquigarrow}$B\\ \hline
  5 & A$\stackrel{p}{\rightsquigarrow}$C\\ \hline
  6 & B$\stackrel{p}{\rightsquigarrow}$E\\ \hline
  8 & D$\stackrel{p}{\rightsquigarrow}$E\\ \hline
 \end{tabular}
 \\
 \\
 \begin{tabular}{l}
  extract-min: D$\stackrel{p}{\rightsquigarrow}$F\\
  path weight: $w(p) = 3$\\
  path reject: \verb@FALSE@\\
 \end{tabular}\\
\end{tabular}
\\
\end{tabular}
\end{center}
\end{figure}
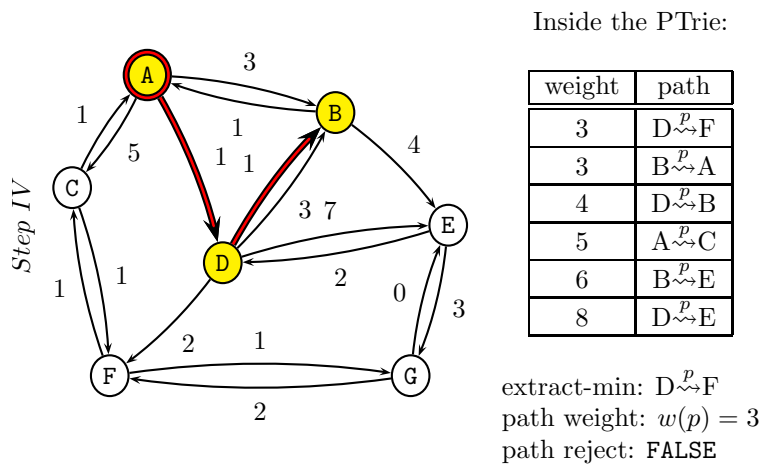
\begin{figure}[!]
\caption{Algorithm compute SDSP of a work, step V\label{SDSPtree5}}
\begin{center}
\begin{tabular}{ p{6.3cm} @{} c}
\begin{tabular}{l}
\begin{pspicture*}(-0.3,-0.2)(6, 5)

\rput[bc](-0.15, 2.5){ \rotateleft{$Step$ $V$} }

\ttfamily
\rput[bc](1.5, 4.5){\ovalnode[doubleline=true,doublecolor=red, fillstyle=solid,fillcolor=yellow]{NodeA}{\textbf{A}}}
\rput[bc](4, 4){\ovalnode[fillstyle=solid,fillcolor=yellow]{NodeB}{\textbf{B}}}
\rput[bc](0.5, 3){\ovalnode{NodeC}{\textbf{C}}}
\rput[bc](2.5, 2){\ovalnode[fillstyle=solid,fillcolor=yellow]{NodeD}{\textbf{D}}}
\rput[bc](5.5, 2.5){\ovalnode{NodeE}{\textbf{E}}}
\rput[bc](1, 0.5){\ovalnode[fillstyle=solid,fillcolor=yellow]{NodeF}{\textbf{F}}}
\rput[bc](5, 0.5){\ovalnode{NodeG}{\textbf{G}}}

\ncarc{->}{NodeA}{NodeB}\Aput{$3$}
\ncarc[doubleline=true,doublecolor=red]{->}{NodeA}{NodeD}\Aput{$1$}
\ncarc{->}{NodeA}{NodeC}\Aput{$5$}

\ncarc{->}{NodeB}{NodeA}\Aput{$1$}
\ncarc{->}{NodeB}{NodeE}\Aput{$4$}

\ncarc{->}{NodeC}{NodeA}\Aput{$1$}
\ncarc{->}{NodeC}{NodeF}\Aput{$1$}

\ncarc[doubleline=true,doublecolor=red]{->}{NodeD}{NodeF}\Aput{$2$}
\ncarc[doubleline=true,doublecolor=red]{->}{NodeD}{NodeB}\Aput{$1$}
\ncarc{->}{NodeD}{NodeE}\Aput{$7$}
\ncarc{<-}{NodeB}{NodeD}\Aput{$3$}

\ncarc{->}{NodeE}{NodeD}\Aput{$2$}
\ncarc{->}{NodeE}{NodeG}\Aput{$3$}

\ncarc{->}{NodeG}{NodeE}\Aput{$0$}
\ncarc{->}{NodeG}{NodeF}\Aput{$2$}

\ncarc{->}{NodeF}{NodeC}\Aput{$1$}
\ncarc{->}{NodeF}{NodeG}\Aput{$1$}
\end{pspicture*}
\end{tabular}
&
\begin{tabular}{c}
 Inside the PTrie:\\
 \\
 \begin{tabular}{|c|c|}
  \hline
  weight & path\\
  \hline\hline
  3 & B$\stackrel{p}{\rightsquigarrow}$A\\ \hline
  4 & D$\stackrel{p}{\rightsquigarrow}$B\\ \hline
  4 & F$\stackrel{p}{\rightsquigarrow}$C\\ \hline
  4 & F$\stackrel{p}{\rightsquigarrow}$G\\ \hline
  5 & A$\stackrel{p}{\rightsquigarrow}$C\\ \hline
  6 & B$\stackrel{p}{\rightsquigarrow}$E\\ \hline
  8 & D$\stackrel{p}{\rightsquigarrow}$E\\ \hline
 \end{tabular}
 \\
 \\
 \begin{tabular}{l}
  extract-min: B$\stackrel{p}{\rightsquigarrow}$A\\
  path weight: $w(p) = 3$\\
  path reject: \verb@TRUE@\\
 \end{tabular}\\
\end{tabular}
\\
\end{tabular}
\end{center}
\end{figure}
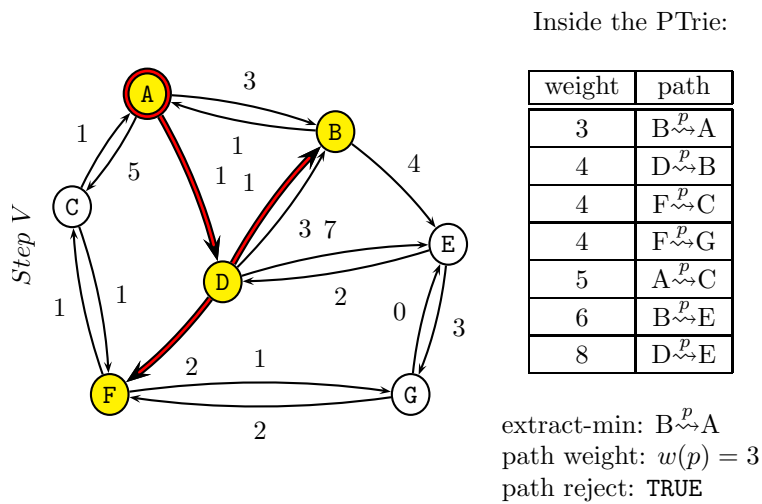
\begin{figure}[!]
\caption{Algorithm compute SDSP of a work, step VI\label{SDSPtree6}}
\begin{center}
\begin{tabular}{ p{6.3cm} @{} c}
\begin{tabular}{l}
\begin{pspicture*}(-0.3,-0.2)(6, 5)

\rput[bc](-0.15, 2.5){ \rotateleft{$Step$ $VI$} }

\ttfamily
\rput[bc](1.5, 4.5){\ovalnode[doubleline=true,doublecolor=red, fillstyle=solid,fillcolor=yellow]{NodeA}{\textbf{A}}}
\rput[bc](4, 4){\ovalnode[fillstyle=solid,fillcolor=yellow]{NodeB}{\textbf{B}}}
\rput[bc](0.5, 3){\ovalnode{NodeC}{\textbf{C}}}
\rput[bc](2.5, 2){\ovalnode[fillstyle=solid,fillcolor=yellow]{NodeD}{\textbf{D}}}
\rput[bc](5.5, 2.5){\ovalnode{NodeE}{\textbf{E}}}
\rput[bc](1, 0.5){\ovalnode[fillstyle=solid,fillcolor=yellow]{NodeF}{\textbf{F}}}
\rput[bc](5, 0.5){\ovalnode{NodeG}{\textbf{G}}}

\ncarc{->}{NodeA}{NodeB}\Aput{$3$}
\ncarc[doubleline=true,doublecolor=red]{->}{NodeA}{NodeD}\Aput{$1$}
\ncarc{->}{NodeA}{NodeC}\Aput{$5$}

\ncarc{->}{NodeB}{NodeA}\Aput{$1$}
\ncarc{->}{NodeB}{NodeE}\Aput{$4$}

\ncarc{->}{NodeC}{NodeA}\Aput{$1$}
\ncarc{->}{NodeC}{NodeF}\Aput{$1$}

\ncarc[doubleline=true,doublecolor=red]{->}{NodeD}{NodeF}\Aput{$2$}
\ncarc[doubleline=true,doublecolor=red]{->}{NodeD}{NodeB}\Aput{$1$}
\ncarc{->}{NodeD}{NodeE}\Aput{$7$}
\ncarc{<-}{NodeB}{NodeD}\Aput{$3$}

\ncarc{->}{NodeE}{NodeD}\Aput{$2$}
\ncarc{->}{NodeE}{NodeG}\Aput{$3$}

\ncarc{->}{NodeG}{NodeE}\Aput{$0$}
\ncarc{->}{NodeG}{NodeF}\Aput{$2$}

\ncarc{->}{NodeF}{NodeC}\Aput{$1$}
\ncarc{->}{NodeF}{NodeG}\Aput{$1$}
\end{pspicture*}
\end{tabular}
&
\begin{tabular}{c}
 Inside the PTrie:\\
 \\
 \begin{tabular}{|c|c|}
  \hline
  weight & path\\
  \hline\hline
  4 & D$\stackrel{p}{\rightsquigarrow}$B\\ \hline
  4 & F$\stackrel{p}{\rightsquigarrow}$C\\ \hline
  4 & F$\stackrel{p}{\rightsquigarrow}$G\\ \hline
  5 & A$\stackrel{p}{\rightsquigarrow}$C\\ \hline
  6 & B$\stackrel{p}{\rightsquigarrow}$E\\ \hline
  8 & D$\stackrel{p}{\rightsquigarrow}$E\\ \hline
 \end{tabular}
 \\
 \\
 \begin{tabular}{l}
  extract-min: D$\stackrel{p}{\rightsquigarrow}$B\\
  path weight: $w(p) = 4$\\
  path reject: \verb@TRUE@\\
 \end{tabular}\\
\end{tabular}
\\
\end{tabular}
\end{center}
\end{figure}
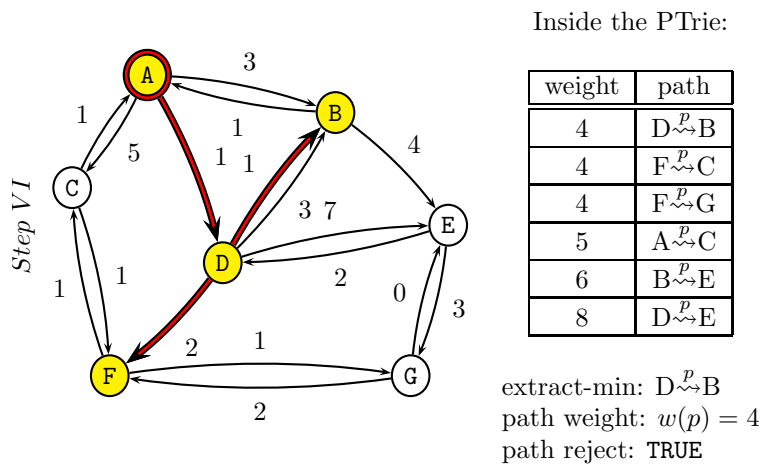
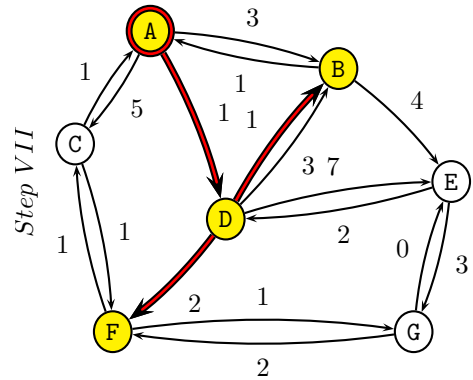
\begin{figure}[!]
\caption{Algorithm compute SDSP of a work, step VII\label{SDSPtree7}}
\begin{center}
\begin{tabular}{ p{6.3cm} @{} c}
\begin{tabular}{l}
\begin{pspicture*}(-0.3,-0.2)(6, 5)

\rput[bc](-0.15, 2.5){ \rotateleft{$Step$ $VII$} }

\ttfamily
\rput[bc](1.5, 4.5){\ovalnode[doubleline=true,doublecolor=red, fillstyle=solid,fillcolor=yellow]{NodeA}{\textbf{A}}}
\rput[bc](4, 4){\ovalnode[fillstyle=solid,fillcolor=yellow]{NodeB}{\textbf{B}}}
\rput[bc](0.5, 3){\ovalnode{NodeC}{\textbf{C}}}
\rput[bc](2.5, 2){\ovalnode[fillstyle=solid,fillcolor=yellow]{NodeD}{\textbf{D}}}
\rput[bc](5.5, 2.5){\ovalnode{NodeE}{\textbf{E}}}
\rput[bc](1, 0.5){\ovalnode[fillstyle=solid,fillcolor=yellow]{NodeF}{\textbf{F}}}
\rput[bc](5, 0.5){\ovalnode{NodeG}{\textbf{G}}}

\ncarc{->}{NodeA}{NodeB}\Aput{$3$}
\ncarc[doubleline=true,doublecolor=red]{->}{NodeA}{NodeD}\Aput{$1$}
\ncarc{->}{NodeA}{NodeC}\Aput{$5$}

\ncarc{->}{NodeB}{NodeA}\Aput{$1$}
\ncarc{->}{NodeB}{NodeE}\Aput{$4$}

\ncarc{->}{NodeC}{NodeA}\Aput{$1$}
\ncarc{->}{NodeC}{NodeF}\Aput{$1$}

\ncarc[doubleline=true,doublecolor=red]{->}{NodeD}{NodeF}\Aput{$2$}
\ncarc[doubleline=true,doublecolor=red]{->}{NodeD}{NodeB}\Aput{$1$}
\ncarc{->}{NodeD}{NodeE}\Aput{$7$}
\ncarc{<-}{NodeB}{NodeD}\Aput{$3$}

\ncarc{->}{NodeE}{NodeD}\Aput{$2$}
\ncarc{->}{NodeE}{NodeG}\Aput{$3$}

\ncarc{->}{NodeG}{NodeE}\Aput{$0$}
\ncarc{->}{NodeG}{NodeF}\Aput{$2$}

\ncarc{->}{NodeF}{NodeC}\Aput{$1$}
\ncarc{->}{NodeF}{NodeG}\Aput{$1$}
\end{pspicture*}
\end{tabular}
&
\begin{tabular}{c}
 Inside the PTrie:\\
 \\
 \begin{tabular}{|c|c|}
  \hline
  weight & path\\
  \hline\hline
  4 & F$\stackrel{p}{\rightsquigarrow}$C\\ \hline
  4 & F$\stackrel{p}{\rightsquigarrow}$G\\ \hline
  5 & A$\stackrel{p}{\rightsquigarrow}$C\\ \hline
  6 & B$\stackrel{p}{\rightsquigarrow}$E\\ \hline
  8 & D$\stackrel{p}{\rightsquigarrow}$E\\ \hline
 \end{tabular}
 \\
 \\
 \begin{tabular}{l}
  extract-min: F$\stackrel{p}{\rightsquigarrow}$C\\
  path weight: $w(p) = 4$\\
  path reject: \verb@FALSE@\\
 \end{tabular}\\
\end{tabular}
\\
\end{tabular}
\end{center}
\end{figure}
\begin{figure}[!]
\caption{Algorithm compute SDSP of a work, step VIII\label{SDSPtree8}}
\begin{center}
\begin{tabular}{ p{6.3cm} @{} c}
\begin{tabular}{l}
\begin{pspicture*}(-0.3,-0.2)(6, 5)

\rput[bc](-0.15, 2.5){ \rotateleft{$Step$ $VIII$} }

\ttfamily
\rput[bc](1.5, 4.5){\ovalnode[doubleline=true,doublecolor=red, fillstyle=solid,fillcolor=yellow]{NodeA}{\textbf{A}}}
\rput[bc](4, 4){\ovalnode[fillstyle=solid,fillcolor=yellow]{NodeB}{\textbf{B}}}
\rput[bc](0.5, 3){\ovalnode[fillstyle=solid,fillcolor=yellow]{NodeC}{\textbf{C}}}
\rput[bc](2.5, 2){\ovalnode[fillstyle=solid,fillcolor=yellow]{NodeD}{\textbf{D}}}
\rput[bc](5.5, 2.5){\ovalnode{NodeE}{\textbf{E}}}
\rput[bc](1, 0.5){\ovalnode[fillstyle=solid,fillcolor=yellow]{NodeF}{\textbf{F}}}
\rput[bc](5, 0.5){\ovalnode{NodeG}{\textbf{G}}}

\ncarc{->}{NodeA}{NodeB}\Aput{$3$}
\ncarc[doubleline=true,doublecolor=red]{->}{NodeA}{NodeD}\Aput{$1$}
\ncarc{->}{NodeA}{NodeC}\Aput{$5$}

\ncarc{->}{NodeB}{NodeA}\Aput{$1$}
\ncarc{->}{NodeB}{NodeE}\Aput{$4$}

\ncarc{->}{NodeC}{NodeA}\Aput{$1$}
\ncarc{->}{NodeC}{NodeF}\Aput{$1$}

\ncarc[doubleline=true,doublecolor=red]{->}{NodeD}{NodeF}\Aput{$2$}
\ncarc[doubleline=true,doublecolor=red]{->}{NodeD}{NodeB}\Aput{$1$}
\ncarc{->}{NodeD}{NodeE}\Aput{$7$}
\ncarc{<-}{NodeB}{NodeD}\Aput{$3$}

\ncarc{->}{NodeE}{NodeD}\Aput{$2$}
\ncarc{->}{NodeE}{NodeG}\Aput{$3$}

\ncarc{->}{NodeG}{NodeE}\Aput{$0$}
\ncarc{->}{NodeG}{NodeF}\Aput{$2$}

\ncarc[doubleline=true,doublecolor=red]{->}{NodeF}{NodeC}\Aput{$1$}
\ncarc{->}{NodeF}{NodeG}\Aput{$1$}
\end{pspicture*}
\end{tabular}
&
\begin{tabular}{c}
 Inside the PTrie:\\
 \\
 \begin{tabular}{|c|c|}
  \hline
  weight & path\\
  \hline\hline
  4 & F$\stackrel{p}{\rightsquigarrow}$G\\ \hline
  5 & A$\stackrel{p}{\rightsquigarrow}$C\\ \hline
  5 & C$\stackrel{p}{\rightsquigarrow}$A\\ \hline
  5 & C$\stackrel{p}{\rightsquigarrow}$F\\ \hline
  6 & B$\stackrel{p}{\rightsquigarrow}$E\\ \hline
  8 & D$\stackrel{p}{\rightsquigarrow}$E\\ \hline
 \end{tabular}
 \\
 \\
 \begin{tabular}{l}
  extract-min: F$\stackrel{p}{\rightsquigarrow}$G\\
  path weight: $w(p) = 4$\\
  path reject: \verb@FALSE@\\
 \end{tabular}\\
\end{tabular}
\\
\end{tabular}
\end{center}
\end{figure}
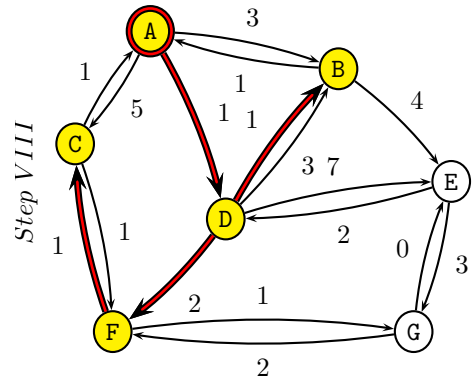
\begin{figure}[!]
\caption{Algorithm compute SDSP of a work, step IX\label{SDSPtree9}}
\begin{center}
\begin{tabular}{ p{6.3cm} @{} c}
\begin{tabular}{l}
\begin{pspicture*}(-0.3,-0.2)(6, 5)

\rput[bc](-0.15, 2.5){ \rotateleft{$Step$ $IX$} }

\ttfamily
\rput[bc](1.5, 4.5){\ovalnode[doubleline=true,doublecolor=red, fillstyle=solid,fillcolor=yellow]{NodeA}{\textbf{A}}}
\rput[bc](4, 4){\ovalnode[fillstyle=solid,fillcolor=yellow]{NodeB}{\textbf{B}}}
\rput[bc](0.5, 3){\ovalnode[fillstyle=solid,fillcolor=yellow]{NodeC}{\textbf{C}}}
\rput[bc](2.5, 2){\ovalnode[fillstyle=solid,fillcolor=yellow]{NodeD}{\textbf{D}}}
\rput[bc](5.5, 2.5){\ovalnode{NodeE}{\textbf{E}}}
\rput[bc](1, 0.5){\ovalnode[fillstyle=solid,fillcolor=yellow]{NodeF}{\textbf{F}}}
\rput[bc](5, 0.5){\ovalnode[fillstyle=solid,fillcolor=yellow]{NodeG}{\textbf{G}}}

\ncarc{->}{NodeA}{NodeB}\Aput{$3$}
\ncarc[doubleline=true,doublecolor=red]{->}{NodeA}{NodeD}\Aput{$1$}
\ncarc{->}{NodeA}{NodeC}\Aput{$5$}

\ncarc{->}{NodeB}{NodeA}\Aput{$1$}
\ncarc{->}{NodeB}{NodeE}\Aput{$4$}

\ncarc{->}{NodeC}{NodeA}\Aput{$1$}
\ncarc{->}{NodeC}{NodeF}\Aput{$1$}

\ncarc[doubleline=true,doublecolor=red]{->}{NodeD}{NodeF}\Aput{$2$}
\ncarc[doubleline=true,doublecolor=red]{->}{NodeD}{NodeB}\Aput{$1$}
\ncarc{->}{NodeD}{NodeE}\Aput{$7$}
\ncarc{<-}{NodeB}{NodeD}\Aput{$3$}

\ncarc{->}{NodeE}{NodeD}\Aput{$2$}
\ncarc{->}{NodeE}{NodeG}\Aput{$3$}

\ncarc{->}{NodeG}{NodeE}\Aput{$0$}
\ncarc{->}{NodeG}{NodeF}\Aput{$2$}

\ncarc[doubleline=true,doublecolor=red]{->}{NodeF}{NodeC}\Aput{$1$}
\ncarc[doubleline=true,doublecolor=red]{->}{NodeF}{NodeG}\Aput{$1$}
\end{pspicture*}
\end{tabular}
&
\begin{tabular}{c}
 Inside the PTrie:\\
 \\
 \begin{tabular}{|c|c|}
  \hline
  weight & path\\
  \hline\hline
  4 & G$\stackrel{p}{\rightsquigarrow}$E\\ \hline
  5 & A$\stackrel{p}{\rightsquigarrow}$C\\ \hline
  5 & C$\stackrel{p}{\rightsquigarrow}$A\\ \hline
  5 & C$\stackrel{p}{\rightsquigarrow}$F\\ \hline
  6 & B$\stackrel{p}{\rightsquigarrow}$E\\ \hline
  8 & D$\stackrel{p}{\rightsquigarrow}$E\\ \hline
 \end{tabular}
 \\
 \\
 \begin{tabular}{l}
  extract-min: G$\stackrel{p}{\rightsquigarrow}$E\\
  path weight: $w(p) = 4$\\
  path reject: \verb@FALSE@\\
 \end{tabular}\\
\end{tabular}
\\
\end{tabular}
\end{center}
\end{figure}
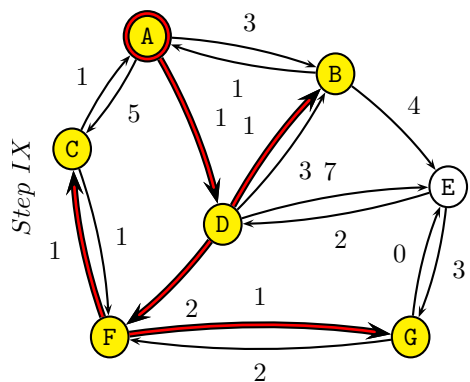
\begin{figure}[!]
\caption{Algorithm compute SDSP of a work, step X\label{SDSPtree10}}
\begin{center}
\begin{tabular}{ p{6.3cm} @{} c}
\begin{tabular}{l}
\begin{pspicture*}(-0.3,-0.2)(6, 5)

\rput[bc](-0.15, 2.5){ \rotateleft{$Step$ $X$} }

\ttfamily
\rput[bc](1.5, 4.5){\ovalnode[doubleline=true,doublecolor=red, fillstyle=solid,fillcolor=yellow]{NodeA}{\textbf{A}}}
\rput[bc](4, 4){\ovalnode[fillstyle=solid,fillcolor=yellow]{NodeB}{\textbf{B}}}
\rput[bc](0.5, 3){\ovalnode[fillstyle=solid,fillcolor=yellow]{NodeC}{\textbf{C}}}
\rput[bc](2.5, 2){\ovalnode[fillstyle=solid,fillcolor=yellow]{NodeD}{\textbf{D}}}
\rput[bc](5.5, 2.5){\ovalnode[fillstyle=solid,fillcolor=yellow]{NodeE}{\textbf{E}}}
\rput[bc](1, 0.5){\ovalnode[fillstyle=solid,fillcolor=yellow]{NodeF}{\textbf{F}}}
\rput[bc](5, 0.5){\ovalnode[fillstyle=solid,fillcolor=yellow]{NodeG}{\textbf{G}}}

\ncarc{->}{NodeA}{NodeB}\Aput{$3$}
\ncarc[doubleline=true,doublecolor=red]{->}{NodeA}{NodeD}\Aput{$1$}
\ncarc{->}{NodeA}{NodeC}\Aput{$5$}

\ncarc{->}{NodeB}{NodeA}\Aput{$1$}
\ncarc{->}{NodeB}{NodeE}\Aput{$4$}

\ncarc{->}{NodeC}{NodeA}\Aput{$1$}
\ncarc{->}{NodeC}{NodeF}\Aput{$1$}

\ncarc[doubleline=true,doublecolor=red]{->}{NodeD}{NodeF}\Aput{$2$}
\ncarc[doubleline=true,doublecolor=red]{->}{NodeD}{NodeB}\Aput{$1$}
\ncarc{->}{NodeD}{NodeE}\Aput{$7$}
\ncarc{<-}{NodeB}{NodeD}\Aput{$3$}

\ncarc{->}{NodeE}{NodeD}\Aput{$2$}
\ncarc{->}{NodeE}{NodeG}\Aput{$3$}

\ncarc[doubleline=true,doublecolor=red]{->}{NodeG}{NodeE}\Aput{$0$}
\ncarc{->}{NodeG}{NodeF}\Aput{$2$}

\ncarc[doubleline=true,doublecolor=red]{->}{NodeF}{NodeC}\Aput{$1$}
\ncarc[doubleline=true,doublecolor=red]{->}{NodeF}{NodeG}\Aput{$1$}
\end{pspicture*}
\end{tabular}
&
\begin{tabular}{c}
 Inside the PTrie:\\
 \\
 \begin{tabular}{|c|c|}
  \hline
  weight & path\\
  \hline\hline
  5 & A$\stackrel{p}{\rightsquigarrow}$C\\ \hline
  5 & C$\stackrel{p}{\rightsquigarrow}$A\\ \hline
  5 & C$\stackrel{p}{\rightsquigarrow}$F\\ \hline
  6 & B$\stackrel{p}{\rightsquigarrow}$E\\ \hline
  6 & E$\stackrel{p}{\rightsquigarrow}$D\\ \hline
  7 & E$\stackrel{p}{\rightsquigarrow}$G\\ \hline
  8 & D$\stackrel{p}{\rightsquigarrow}$E\\ \hline
 \end{tabular}
 \\
 \\
 \begin{tabular}{l}
  extract-min: remaining \\
  path weight:\\
  path reject: \verb@TRUE@\\
 \end{tabular}\\
\end{tabular}
\\
\end{tabular}
\end{center}
\end{figure}
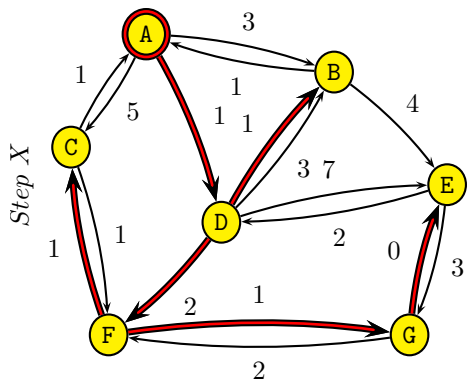
\begin{figure}[!]
\caption{Algorithm compute SDSP of a work: SDSP tree\label{SDSPtree11}}
\begin{center}
\begin{pspicture*}(0.0,-0.2)(6, 5)
\rput[bc](2.7,4.85){\rnode{ROOT}{root}}
\ttfamily
\rput[bc](1.5, 4.5){\ovalnode[doubleline=true,doublecolor=red, fillstyle=solid,fillcolor=yellow]{NodeA}{\textbf{A}}}
\rput[bc](4, 4){\ovalnode[fillstyle=solid,fillcolor=yellow]{NodeB}{\textbf{B}}}
\rput[bc](0.5, 3){\ovalnode[fillstyle=solid,fillcolor=yellow]{NodeC}{\textbf{C}}}
\rput[bc](2.5, 2){\ovalnode[fillstyle=solid,fillcolor=yellow]{NodeD}{\textbf{D}}}
\rput[bc](5.5, 2.5){\ovalnode[fillstyle=solid,fillcolor=yellow]{NodeE}{\textbf{E}}}
\rput[bc](1, 0.5){\ovalnode[fillstyle=solid,fillcolor=yellow]{NodeF}{\textbf{F}}}
\rput[bc](5, 0.5){\ovalnode[fillstyle=solid,fillcolor=yellow]{NodeG}{\textbf{G}}}
\rput[bc](0.3,4.85){\rnode{NIL}{NIL}}

\ncarc[doubleline=true,doublecolor=gray]{<-}{NodeA}{NodeD}\Aput{$1$}
\ncarc[doubleline=true,doublecolor=gray]{<-}{NodeD}{NodeF}\Aput{$2$}
\ncarc[doubleline=true,doublecolor=gray]{<-}{NodeD}{NodeB}\Aput{$1$}
\ncarc[doubleline=true,doublecolor=gray]{<-}{NodeG}{NodeE}\Aput{$0$}
\ncarc[doubleline=true,doublecolor=gray]{<-}{NodeF}{NodeC}\Aput{$1$}
\ncarc[doubleline=true,doublecolor=gray]{<-}{NodeF}{NodeG}\Aput{$1$}
\ncarc[doubleline=true,doublecolor=gray]{<-}{NIL}{NodeA}
\pcline{<-}(1.9, 4.65)(3.3, 4.65)

\end{pspicture*}
\\
Legend: gray arcs denote pointer `back' on vertices\\
\verb@       @ yellow vertex with red aureola is a root\\
\verb@       @ yellow vertices without red aureola are leaves
\end{center}
\end{figure}
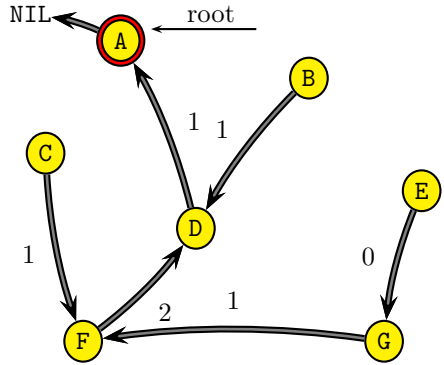

\subsubsection{Analysis of correctness}
The algorithm shown here starts to analyze the graph and create the shortest 
paths to the source vertex. If the graph $G$ is not strongly connected%
~\footnote{%
A directed graph is strongly connected if every two vertices are reachable 
from each other.%
} 
the algorithm which solves SDSP problem and starts its work from the source 
vertex will calculate SDSP tree of  connected components%
~\footnote{%
The connected components of a graph are the equivalence classes of vertices 
under the ``is reachable from'' relation.
} 
containing the source vertex; $G_{s \in V} = (s, V, E)$.

The algorithm does not use weight relaxation. Arcs added to the SDSP tree 
are not modified any more. Only the order of taking the paths out of the 
PTrie determines the choice of arcs or paths, starting from the 
minimum-weight path. It's worth remembering, however, that the weight of 
each vertex which has not been added to SDSP tree should be increased by 
the weight of the path which brought us there before inserting it to PTrie. 
So the vertices to which lead the minimum-weight path are visited always 
but only once.
The SDSP tree is represented by the `$back$' connected with the vertices. 
That's means that for arbitrary graph $G=(V,E)$ with the directed source 
vertex the algorithm define the tree,  which is the subgraph of the 
predecessor of the graph $G$ as graph $T_{back} = (V, E_{back})$.
Therefore the algorithm is correct because the shortest paths are composed 
of shortest paths. The proof of this is based on the notion that if there 
was a shorter path than any sub-path, then the shorter path should replace 
that sub-path to make the whole path shorter. That's why the subgraph of 
predecessors $T_{back}$ is the Shortest Path Tree.\break
\subsubsection{Analysis of the algorithm bound}
Body loop which inserts arcs to PTrie is $\Theta(|E|)$ time cost. 
The operation of PTrie for constant length (size of) type weight of arc are 
$\Theta(\frac{Mconst}{K} + K) = O(1)$. To look through each of vertex graph 
the algorithm require $\Theta(|V|)$ time. Therefore worst-case time 
complexity equals $\Theta(|V| + |E|(\frac{Mconst}{K} + K)) = O(|V| + |E|)$ 
time.\break
\subsubsection{A simple example of the use of algorithm computing SDSP problem in C++}

The algorithm builds SDSP tree for graph presented in 
{\normalfont{[Figure~\ref{SDSPtree1}]}}.
%
\\
\lstset{language=C++, flexiblecolumns=true}
%
%
\rule[-1ex]{\textwidth}{2pt}
\begin{small}
\begin{lstlisting}
#include<iostream>

// The Report contains the PTrie source code:
// http://techreports.library.cornell.edu:8081/
// Dienst/UI/1.0/Display/cul.cis/TR2006-2023
//
// Or, the SourceForge.net project: 
// http://ptrie.sourceforge.net
#include "PTrie.hpp" /*
                     ** http://ptrie.sourceforge.net/source/PTrie.hpp
                     */


// <begin> namespace dplaneta
namespace dplaneta{

struct Vertex;
struct Neighbors;
struct Arc;

/*******************************************************************************
** The definition of the structure which represents vertex.
*/
struct Vertex{
       char data; // Label
       Neighbors *list; //Adjacent List
       
       Vertex *back;
       unsigned backWeight; 
       
       // The metod is responsible for  connecting vertices with each other.
       Vertex& attach(Vertex *attach, unsigned weight);
       
       Vertex(char name); 
       ~Vertex(void);
       };

/*******************************************************************************
** The definition of the structure responsible for the organization 
** of adjacent list.
*/
struct Neighbors{
       Neighbors *next; // Adjacent List [the singly-linked list]
       
       // Arc representation
       Vertex *link; 
       unsigned weight;
       
       Neighbors(Vertex *l, unsigned w, Neighbors *n);
       ~Neighbors(void);
       };
       




Vertex::Vertex(char name): list(0), back(0), data(name){}
Vertex::~Vertex(void){
        delete list;
        back = 0;
        list = 0; 
        }


// The method 'attach' implemented in vertex structure is responsible for the 
// connection of the vertices. For example, to connect vertex 'A' with 'B'
// [Vertex *a = new Vertex('A'), *b = new Vertex('B')]
// We add the edge which connects vertex A to its adjacent list.
// [a->add(b, lenght);]
// Because of this we get the arc (directed edge) joining 'A' and 'B'.
// To simulate the edge between 'A' and 'B', also vertex 'B' should contain the 
// arc of the same length which joins it with 'A'. [b->add(a, lenght);]
// (A) <===> (B).
Vertex& Vertex::attach(Vertex *attach, unsigned weight){
        list = new Neighbors(attach, weight, list);
        return *this;
}       

       
   
   
       
Neighbors::Neighbors(Vertex *l, unsigned w, Neighbors *n): 
           link(l), weight(w), next(n){}

Neighbors::~Neighbors(void){
                            delete next;
                            next = 0;
                            }




/*******************************************************************************
** The definition of the structure which represents arc is necessary 
** inside the PTrie. The structure must have the implemented overloaded 
** operators used by PTrie.
*/
struct Arc{ 
       unsigned weight, pathWeight;
       Vertex *tail, *head;
       unsigned operator>>(const unsigned i) const 
            { return (pathWeight>>i); }
            
       bool operator!=(const Arc& obj) const 
            { return this->pathWeight!=obj.pathWeight; }
            
       bool operator==(const Arc& obj) const 
            { return this->pathWeight==obj.pathWeight; }
       };





/*******************************************************************************
** Supportive function used by PTrie to return size of Type<Arc.weight>
*/
inline size_t sizeFunc(const class Arc& obj){
       return sizeof(obj.weight);
}




// If You want to see how the arcs are analyzed by the algorithm...
// #define TEST

/*******************************************************************************
** The algorithm builds the SDSP tree by proper arrangement of support 
** varieties 'back' which are located in every vertex.
*/
void SDSP(Vertex* sourceVertex){
     register Arc temp;
     register const Arc *p;
     register const Neighbors *t;
     PTrie<Arc> Q(sizeFunc);
     PTrie<Arc>::iterator iter=Q;
     
#ifdef TEST
std::cout<<"Insert the adjacent list of the source vertex to PTrie:\n";
#endif
     for(t = sourceVertex->list; t!=NULL; t=t->next){
         temp.weight = temp.pathWeight = t->weight;
         temp.tail = sourceVertex;
         temp.head = t->link;
         Q.insert(temp);
#ifdef TEST
std::cout<<"["<<sourceVertex->data<<"]--("<<t->weight<<")->["<<t->link->data<<"]\n";
#endif
     }

#ifdef TEST     
std::cout<<std::endl;
#endif
     
     // All arcs accessible from the source vertex are analyzed.
     while(p = Q.minimum()){
        
#ifdef TEST
std::cout<<"\tCame from the ["<<p->tail->data<<"]; State of the PTrie: \n";
iter.begin();
while(iter){
  std::cout<<"\t["<<(*iter).tail->data<<"]--("<<(*iter).pathWeight<<")->["
           <<(*iter).head->data<<"]\n";
  iter++;
}
std::cout<<std::endl;
#endif
        
        if(p->head->back==NULL && p->head!=sourceVertex){
        
#ifdef TEST   
std::cout<<"\nThe vertex is chosen: ["<<p->head->data
         <<"] and insert the adjacent list to PTrie:\n";
#endif
           
           ((Arc*)p)->head->back = p->tail;
           p->head->backWeight = p->weight;
           
           // Insert the adjacent list of the current vertex to PTrie
           for(t = p->head->list; t!=NULL; t=t->next){
              temp.weight = t->weight;
              temp.pathWeight = t->weight +  p->pathWeight;
              temp.tail = p->head;
              temp.head = t->link;
              Q.insert(temp);
              
#ifdef TEST
std::cout<<"["<<temp.tail->data<<"]--("<<temp.weight<<")->["
         <<temp.head->data<<"]"\
" weight of path = "<<temp.pathWeight<<std::endl;
#endif
           }

#ifdef TEST             
std::cout<<std::endl; 
#endif              
        }
        Q.remove(*p);
        
     }
     
     
}





// The function walks on the path from the choosen vertex to the source vertex.
void Walk(const Vertex *v){
     while(v){
              std::cout<<"["<<v->data<<"]";
              if(v->back) std::cout<<"--("<<v->backWeight<<")->";
              v = v->back;
     }
     std::cout<<std::endl;
       
}





// The function shows the adjacent list of the choosen vertex.
void ShowAdjecentLists(const Vertex *v){
     for(const Neighbors *t = v->list; t!=NULL; t=t->next)
        std::cout<<"\t["<<v->data<<"]--("<<t->weight<<")->["
                 <<t->link->data<<"]\n";
     std::cout<<std::endl;
}


}// <end> namespace dplaneta





/*******************************************************************************
** This example source code demonstrates how you can use a shown algorithm 
** computing SDSP problem.
*/
int main(int argc, char *argv[]){
// We create the graph vertices 
dplaneta::Vertex a('A'), b('B'), c('C'), d('D'), e('E'), f('F'), g('G');
dplaneta::Vertex *sourceVertex;
        
// Connecting the graph vertices.
a.attach(&b, 3).attach(&c, 5).attach(&d, 1);
b.attach(&a, 1).attach(&e, 4);
c.attach(&f, 1).attach(&a, 1);
d.attach(&b, 1).attach(&b, 3).attach(&e, 7).attach(&f, 2);
e.attach(&d, 2).attach(&g, 3);
f.attach(&g, 1).attach(&c, 1);
g.attach(&e, 0);
             
std::cout<<"Show adjecent lists:\n-----------------------\n";
std::cout<<"A:\n"; ShowAdjecentLists(&a);
std::cout<<"B:\n"; ShowAdjecentLists(&b);
std::cout<<"C:\n"; ShowAdjecentLists(&c);
std::cout<<"D:\n"; ShowAdjecentLists(&d);
std::cout<<"E:\n"; ShowAdjecentLists(&e);
std::cout<<"F:\n"; ShowAdjecentLists(&f);
std::cout<<"G:\n"; ShowAdjecentLists(&g);
std::cout<<"-----------------------\n\n";

// We choose on arbitral vertex from which the algorithm will build SDSP tree.
sourceVertex = &a;
       

// We create the SDSP tree, by suitably setting supportive 
// variable 'back' which are located in each vertex.
dplaneta::SDSP(sourceVertex);
       
// We check the  paths.
std::cout<<"SDSP Path(es?):\n";
dplaneta::Walk(&a);
dplaneta::Walk(&b);
dplaneta::Walk(&c);
dplaneta::Walk(&d);
dplaneta::Walk(&e);
dplaneta::Walk(&f);
dplaneta::Walk(&g);

return 0;
}
\end{lstlisting}
\end{small}
\rule[-1ex]{\textwidth}{2pt}
\section{Conclusions}
I have shown linear worst-worst case algorithms based on PTrie which compute 
the basic Network problems. Despite the fact that PTrie is based on 
digital data, it can be used to store positive integer, integer but also 
real numbers. Because all quantities in computer are represented by binary 
words. That's why the weight of arc can be defined not only by positive 
integer, but also by real number, or even by string. Thanks PTrie, which 
is stable, during computing the MST, SSSP and SDSP problems, we not only 
focus on the Shortest Path in relation to weight of arcs, but also to the 
amount of vertices and arcs on the path too. Time complexity of mentioned 
algorithms equals $O(|V| + |E|)$. Memory bound of algorithms equals memory 
bound of PTrie. PTrie memory bound equals $\Theta(\frac{log_{2^K}N(2^{K+1})}{K})$.
Presented algorithms not only get the fastest asymptotic running time, but 
they are also very practicable and can be easily implemented.

%
%
%
\end{document}